\documentclass[manuscript,revised]{geophysics}
\usepackage{amsmath}
\usepackage{newtxmath}
\usepackage{booktabs}
\usepackage{subfloat}
\usepackage{hyperref}
\usepackage{multirow}
\usepackage{float}
\usepackage{bm}
\begin{document}  

\title{Self-Supervised Diffusion Model for 3-D Seismic Data Reconstruction}
\renewcommand{\thefootnote}{\fnsymbol{footnote}}
\address{
	\footnotemark[1] Department of Communication Engineering\\
Northeast Electric Power University\\
	Jilin, Jilin 132012, China\\
	\footnotemark[2] College of Instrumentation and Electrical Engineering,\\
	Jilin University \\
	Changchun, Jilin 130026, China (E-mail: 18186829038@163.com).\\	
}

\author{Xinyang Wang\footnotemark[1], Qianyu Ge\footnotemark[1], Xintong Dong\footnotemark[2], Shiqi Dong\footnotemark[1], Tie Zhong\footnotemark[1]}

%\footer{GEO-2015}
\lefthead{Dong etc.}
\righthead{S2DM: 3D Reconstruction}
\maketitle

\maketitle	

\begin{abstract}
	Seismic data reconstruction is an effective tool for compensating nonuniform and incomplete seismic geometry. Compared with methods for 2D seismic data, 3D reconstruction methods could consider more spatial structure correlation in seismic data. In the early studies, 3D reconstruction methods are mainly theory-driven and have some limitations due to their prior assumptions on the seismic data. To release these limitations, deep learning-based reconstruction methods rise and show potential in dealing with reconstruction problems. However, there are mainly two shortcomings in existing deep learning-methods. On the one hand, most of existing deep learning-based methods adopt the convolutional neural network, having some difficulties in dealing with data with complex or time-varying distributions. Recently, the diffusion model has been reported to possess the capability to solve data with complex distributions by gradually complicating the distribution of data to optimize the network. On the other hand, existing methods need enough paired-data to train the network, which are very hard to obtain especially for the starved 3D seismic data. Deep prior-based unsupervised and sampling-based self-supervised networks offer an available solution to this problem. In this paper, we develop a self-supervised diffusion model (S2DM) for 3D seismic data reconstruction. The proposed model mainly contains a diffusion restoration model and a variational time-spatial module. Extensive synthetic and field experiments demonstrate the superiority of the proposed S2DM algorithm.
\end{abstract}

\section*{Introduction}

Complete and dense seismic data are conductive to the downstream processing, inversion, and interpretation tasks \cite[]{1,2,3,36,35}. However, seismic data are commonly incomplete and missing traces are likely to occur in some field seismic data due to the economy, acquisition environment, and geographical factors. Thus, reconstructing missing traces is an indispensable step in seismic data processing. 

During the past several decades, many algorithms for reconstructing missing traces in seismic data have been put forward and developed \cite[]{4,5,6,7}. Most of these algorithms mainly focus on processing the 2D frequency slice \cite[]{4}, 2D time slice or the space slice of 3D data one by one \cite[]{5, 6}, including frequency filtering-based \cite[]{8,7}, wavefield operator-based \cite[]{9}, sparse transform-based \cite[]{11,10,14}, matrix-rank degradation-based \cite[]{12,13}, and deep learning-based reconstruction algorithms \cite[]{6,7}, and so on. These reconstruction algorithms exhibit potential abilities in reconstructing missing traces. However, compared with the algorithms for the 3D seismic data, these algorithms ignore the spatial structure correlation along the third dimension. 

In recent years, 3D reconstructing algorithms have seen significant advancements thanks to the increase in computer power. In the previous research stage, traditional reconstruction methods such as the frequency-space domain predictive filtering methods \cite[]{15,16}, rank reduction methods \cite[]{22,17} and sparse transform-based methods \cite[]{18,19} have been designed and applied in 3D seismic data reconstruction successfully. These methods can tackle the reconstruction problem to some degree, but their performances still can be further improved.

Recently, deep learning-based methods become increasingly popular for tackling the reconstruction problem of 3D data. By establishing a complete dataset, some CNN-based networks \cite[]{21,24,34} and GAN-based networks \cite[]{23,25} are designed and trained in a supervised manner. However, due to the expensive cost, very limited seismic data are available for training. Additionally, these methods might work well in distribution, although they are affected by distribution changes. Thus, supervised deep learning methods might lack the generalization ability in the reconstruction of seismic data.

Aiming at these problems, some researchers attempt to design networks in an unsupervised manner. Unsupervised deep learning methods mine the intrinsic characteristics of the input data without any preset labels. Deep image prior (DIP) is one of the most widely applied unsupervised deep learning algorithms. Some DIP-based algorithms have been designed and developed for seismic data reconstruction \cite[]{26,27}. These methods design the appropriate net architecture and develop the loss function to update the parameters iteratively without any extra labels. As a representative type of unsupervised methods, self-supervised methods take advantage of certain properties of the input data as the supervisory information, guiding the network to learn valuable features. Existing reconstruction methods mainly focus on sampling the incomplete data, combining the sampled data and incomplete data as a training pair \cite[]{28,29,30}.

Motivated by the success of self-supervised learning and the diffusion model, we develop a self-supervised diffusion model (S2DM) for 3D seismic data reconstruction in this article. The main contributions of this paper are as follows:

1) We propose a self-supervised diffusion model (S2DM) for 3D seismic data reconstruction, which integrates the reconstruction task as an inverse problem solved by a denoising diffusion time-spatial model. As far as we know, this is the first article to develop the diffusion model for seismic data processing using a self-supervised approach.

2) In the proposed S2DM, we adopt the self-supervised learning mechanism to train the network without any dependence on the extra data, enhancing the generalization performance.

3) We verify the effectiveness of the proposed S2DM using multiple synthetic and field seismic data. We can reveal from the results that the proposed S2DM is superior to the traditional and other DL-based methods.

In this paper, we first introduce the proposed S2DM for 3D seismic data reconstruction in detail; then the performance of the S2DM is evaluated and compared with some representative methods by using synthetic and field data. Next, we give a discussion on the performance analysis of S2DM. Finally, we conclude this paper.  

\section{Methods}

\subsection{Degradation Model for Seismic Data Reconstruction}
Typically, real observed seismic data are commonly with missing traces. The aim of seismic data reconstruction is to fill in the gaps from the observed incomplete data through a known linear degradation model. Generally speaking, the degradation model for seismic data reconstruction is given as:
\begin{equation}
	\label{eq:1}
	{\bf{y = Hx}},
\end{equation}
where  ${\bf{x}} \in {\mathbb{R}^{m \times n \times p}}$ stands for the complete data,  ${\bf{y}} \in {\mathbb{R}^{m \times n \times p}}$ denotes the observed incomplete data;  ${\bf{H}}$ is the degradation matrix depending on the missing indexes. Therefore, solving $\bf{x}$ from a given observed data $\bf{y}$ becomes a blind ill-posed inverse problem. To solve this inverse problem, we design a self-supervised diffusion model in the following.

\subsection{Diffusion Process for Seismic Data Reconstruction}
Recently, Kawar proposed a denoising diffusion restoration model (DDRM) in \cite{31}, which uses the pre-trained model to solve the linear inverse problem in 2D image restoration. Motivated by the excellent ability in DDRM, we propose S2DM to complete the reconstruction task for 3D seismic data. Compared with DDRM, S2DM not only has the ability to solve the linear problem in 3D data but also trains the network in a self-supervised learning manner without any extra training data to pre-train the network. The proposed S2DM is mainly composed of two key procedures: a forward diffusion process and a reverse diffusion process. Noise is gradually added to the seismic data in the forward diffusion process, transforming the data ${{\bf{x}}_t}$ into Gaussian noise ${x_T}\sim N(0,1)$  over $T$ steps. The diffusion models here are generative models parameterized with a Markov chain conditioned on $\bf{y}$: ${{\bf{x}}_T} \to {{\bf{x}}_{T - 1}} \to  \cdots  \to {{\bf{x}}_1} \to {{\bf{x}}_0}$. In the forward direction, the fixed Markov chain at each step is given by:

\begin{equation}
	\label{eq:2}
	q({{\bf{x}}_{t+1}}|{{\bf{x}}_{t}}) = N({{\bf{x}}_{t+1 }};\sqrt {1 - {\beta _{t+1}}} {{\bf{x}}_{t}},\sqrt {{\beta _{t+1 }}} {\bf{I}})
\end{equation}
Where ${\beta _{t+1}}$ is a hyperparameter that controls the noise level at time step $t+1$; ${{\bf{x}}_{t+1}}$ and ${{\bf{x}}_{t}}$ are the obtained sample at time step $t+1$ and $t$, respectively.

In the reverse diffusion process, we leverage a self-supervised variational time-spatial network to gradually eliminate the noise and reconstruct data ${{\bf{x}}_t}$  from ${{\bf{x}}_{t+1}}$  conditioned on $\bf{y}$ as shown in Figure 1. The details of VTSM and reverse diffusion process are discussed in the following.

\subsection{Variational Time-Spatial Module}
The Variational Time-Spatial Module (VTSM) contains an untrained 3D network and a variational inference-based loss function. The untrained 3D network can leverage the intrinsic structure in the seismic data to eliminate the noise. Meanwhile, the variational inference-based loss promotes the untrained 3D network to learn the posterior probability distribution and recover missing traces from the Gaussian noise. 

a) The untrained 3D time-spatial network

In VTSM, we design a 3D denoising network based on the architecture of 3D-UNet as shown in Figure 2. To extract features from the 3D seismic data, all convolution operations used in this network are based on the 3D kernels. Like other U-shaper networks, the proposed 3D network contains an encoder path and a decoder path. In the encoder path, we first use two CNN blocks consisting of “Conv3D + Batch Normalization (BN)+ReLu” to extract features from the input data at the first level. Then to extract multi-scale features, the 3D maximum pooling operations with the kernel size of $3 \times 3 \times 3$  are applied. From ${2^{{\rm{nd}}}}$  to ${5^{{\rm{th}}}}$  level, we introduce an aggregation block (AB) to enhance the feature extraction ability as shown in Figure 3. In the decoder path, each level contains an up-sampling operation of the feature map in the previous level achieved by a 3D convolution kernel with the size of  $3 \times 3 \times 3$. The gray arrow indicates the concatenation operation between the feature map produced by the encoder and the feature map in the decoder that has been up-sampled at the corresponding level. Once the feature maps are concatenated, an AB block is employed to extract deeper features. At the last layer, a 3D convolution with $1 \times 1 \times 1$ with a $1 \times 1 \times 1$ kernel is adopted to predict the denoising results. 

The details of AB block can be seen in Figure 3. Specifically, the AB block mainly contains four stages. The input of AB block is firstly fed into a point-wise 3D convolution with a size of $1 \times 1 \times 1$ . We employ a $1 \times 1 \times 1$ point-wise 3D convolution and a residual block (RB) as shown in Figure 3 to extract features repeatedly. Subsequently, a 3D convolution with the size of $3 \times 3 \times 3$  is adopted to extract deep features from the output of the RB block and then concatenated with the output of the point-wise convolution. Finally, we use a 3D convolution with the size of  $1 \times 1 \times 1$ to adjust the number of channels.

b)	The variational interference-based loss function

To generate the complete seismic data and update the parameters of the network through the reverse diffusion process, a variational interference-based loss function is established here. 
In the reverse diffusion process, we recover the sample ${\bf{x}}_{t}$ based on the transition probability $q({{\bf{x}}_t}|{{\bf{x}}_{t + 1}},{{\bf{x}}_0},{\bf{y}})$ at the time step $t$. To promote the network to learn the posterior probability distribution, the learnable generative process is denoted as ${p_{{\theta _t}}}({{\bf{x}}_t}|{{\bf{x}}_{t + 1}},{\bf{y}})$ by replacing the  ${{\bf{x}}_0}$ in the transition probability $q({{\bf{x}}_t}|{{\bf{x}}_{t + 1}},{{\bf{x}}_0},{\bf{y}})$  with ${{\bf{x}}_{{\theta _t}}}$ , namely:

\begin{equation}
	\label{eq:3}
	{p_{{\theta _t}}}({{\bf{x}}_t}|{{\bf{x}}_{t + 1}},{\bf{y}}) \buildrel \Delta \over = q({{\bf{x}}_t}|{{\bf{x}}_{t + 1}},{{\bf{x}}_{{\theta _t}}},{\bf{y}})
\end{equation}
where ${\theta _t}$  denotes the parameter set of the network at time step $t$ .

In the reverse diffusion process, the purpose of S2DM is to update the parameters ${\theta _t}$ to make the ${p_{{\theta_t}}}({{\bf{x}}_t}|{{\bf{x}}_{t+1}},{\bf{y}})$ as close to $q({{\bf{x}}_t}|{{\bf{x}}_{t+1}},{{\bf{x}}_0},{\bf{y}})$ by maximizing a variational lower bound:

\begin{equation}
	\label{eq:4}
	\begin{array}{l}
		\mathop {\arg \max }\limits_\theta {\mathbb{E}} {_{q({{\bf{x}}_0}),q({\bf{y}}|{{\bf{x}}_0})}}[\log {p_\theta }({{\bf{x}}_0}|{\bf{y}})]\\
		\mathop { \ge \arg \max }\limits_\theta {\mathbb{E}} {_{q({{\bf{x}}_{0:T}}),q({\bf{y}}|{{\bf{x}}_0})}}[\log \frac{{{p_\theta }({{\bf{x}}_{0:T}}|{\bf{y}})}}{{q({{\bf{x}}_{1:T}}|{{\bf{x}}_0},{\bf{y}})}}]\\
		= \mathop {\arg \max }\limits_\theta {\mathbb{E}} {_{q({{\bf{x}}_{0:T}}),q({\bf{y}}|{{\bf{x}}_0})}}[\log \frac{{{p_\theta }({{\bf{x}}_T}|{\bf{y}})}}{{q({{\bf{x}}_T}|{{\bf{x}}_0},{\bf{y}})}} + \sum\limits_{t = 0}^{T - 1} {\frac{{{p_\theta }({{\bf{x}}_t}|{{\bf{x}}_{t + 1}},{\bf{y}})}}{{q({{\bf{x}}_t}|{{\bf{x}}_{t + 1}},{{\bf{x}}_0},{\bf{y}})}}} ]\\
		= \mathop {\arg \max }\limits_\theta {\mathbb{E}} {_{q({{\bf{x}}_{0:T}}),q({\bf{y}}|{{\bf{x}}_0})}}[\log \frac{{{p_\theta }({{\bf{x}}_T}|{\bf{y}})}}{{q({{\bf{x}}_T}|{{\bf{x}}_0},{\bf{y}})}} + \sum\limits_{t = 1}^{T - 1} {\frac{{{p_\theta }({{\bf{x}}_t}|{{\bf{x}}_{t + 1}},{\bf{y}})}}{{q({{\bf{x}}_t}|{{\bf{x}}_{t + 1}},{{\bf{x}}_0},{\bf{y}})}}}  + \log {p_\theta }({{\bf{x}}_0}|{{\bf{x}}_1},{\bf{y}})]\\
		= \mathop {\arg \max }\limits_\theta {\mathbb{E}} {_{q({{\bf{x}}_{0:T}}),q({\bf{y}}|{{\bf{x}}_0})}}[ - {D_{KL}}({p_\theta }({{\bf{x}}_T}|{\bf{y}})||q({{\bf{x}}_T}|{{\bf{x}}_0},{\bf{y}}))]\\
		+{\mathbb{E}} {_{q({{\bf{x}}_{0:T}}),q({\bf{y}}|{{\bf{x}}_0})}}[\sum\limits_{t = 1}^{T - 1} { - {D_{KL}}({p_\theta }({{\bf{x}}_t}|{{\bf{x}}_{t + 1}},{\bf{y}})||q({{\bf{x}}_t}|{{\bf{x}}_{t + 1}},{{\bf{x}}_0},{\bf{y}}))} ] + \\
		{\mathbb{E}}{_{q({{\bf{x}}_{0:T}}),q({\bf{y}}|{{\bf{x}}_0})}}[\log {p_\theta }({{\bf{x}}_0}|{{\bf{x}}_1},{\bf{y}})]
	\end{array}
\end{equation}

In the forward diffusion process, the sample ${\bf{x}}_t$ can be written as:
\begin{equation}
	\label{eq:5}
	\begin{array}{l}
		{\bf{x}}_{t} = \sqrt{{\bar \alpha _t}}{\bf{x}}_{0}+\sqrt{1-{\bar \alpha _t}}\epsilon 
		
	\end{array}
\end{equation}

where $\epsilon \sim N(0,1)$; ${\bar \alpha _t} = \prod\limits_{i = 1}^t {(1 - {\beta _i})} $.

The variational lower bound can be reduced to the following variational interference-based loss function, see details in \cite[]{31}:

\begin{equation}
	\label{eq:6}
	\mathop {\arg \min }\limits_\theta  ||{{\bf{x}}_t} - \sqrt{\bar \alpha _t}{\textbf{\textit{f}}_\theta }({{\bm{\chi}}})||_F^2
\end{equation}
Where ${\bm{\chi}}$ denotes the 3D Gaussian noise. Given a degraded observation ${{\bf{x}}_{t+1}}$, the parameters $\theta_{t}$ of the network can be updated based on the loss function \eqref{eq:6}. Then, we could receive the estimate of ${{\bf{x}}_{{\theta _{t }}}}$  via  ${{\textbf{\textit{f}}}_{\theta_{t}} }({\bm{\chi}})$, and then ${{\bf{x}}_{t}}$  could be sampled from ${p_{{\theta _{t}}}}({{\bf{x}}_{t }}|{{\bf{x}}_{t+1}},{\bf{y}})$. In this case, the diffusion process can be iteratively reversed in a self-supervised fashion, and generate the estimate of ${{\bf{x}}_0}$  at last. 

\subsection{Reverse Diffusion Process}
Motivated by solving noisy inverse problems stochastically (SNIPS) algorithm in \cite[]{33}, we first perform a singular value decomposition (SVD) of the matrix ${\bf{H}}$, and perform the diffusion process in the SVD domain. The SVD of the matrix ${\bf{H}}$ is denoted as:

\begin{equation}
	\label{eq:7}
	{\bf{H}} = {\bf{U}}{\bf{\Sigma}} {{\bf{V}}^T}
\end{equation}
Where ${\bf{U}} \in {\mathbb{R}^{m \times m}}$  and ${\bf{V}} \in {\mathbb{R}^{n \times n}}$ are orthogonal matrices.  ${\bf{\Sigma}}  \in {\mathbb{R}^{m \times n}}$  is a diagonal matrix with the singular values ${s_1} \ge {s_2} \ge  \cdots {s_n} \ge {s_{n + 1}} =  \cdots  = {s_m} = 0$  if  $n \le m$. The aim of SVD is to associate the noise in the noisy measurement ${\bf{y}}$  with the diffusion noise in  ${{\bf{x}}_{1:T}}$, making the restoration result  ${{\bf{x}}_0}$ depend on the input ${\bf{y}}$. Additionally, we can use SVD to identify the missing areas in the degraded data  ${\bf{y}}$, and then generate the missing traces and denoise the observed data in the reverse diffusion process.

Different from most diffusion-based methods using the pre-trained model \cite[]{31}, S2DM is trained in a self-supervised manner. Since the previous diffusion-based methods optimize the denoising model by using a large amount of training data, they can generate a satisfactory result ${\bf{x}}_0$  from arbitrary  ${{\bf{x}}_t}$, even if  ${{\bf{x}}_t}$ looks like the Gaussian noise. However, since S2DM is trained in a self-supervised manner, a relatively good initial ${{\bf{x}}_{{t_0}}}$  is needed to produce the satisfactory result  ${\bf{x}}_0$, where  ${t_0} < T$.

In the SVD domain, we use the following notations to denote the variables: ${\bar {\bf{x}}_t} = {{\bf{V}}^T}{{\bf{x}}_t}$  and  $\bar {\bf{x}}_t^{(i)}$ denotes the  $i$th index of the vector  ${\bar {\bf{x}}_t}$; ${\bar {\bf{y}}_t} = {{\bf{\Sigma}} ^\dag }{{\bf{U}}^T}{{\bf{y}}_t}$  and $\bar {\bf{y}}_t^{(i)}$  denotes the  $i$th index of the vector  ${\bar {\bf{y}}_t}$, where  $\dag $ is the Moore-Penrose pseudo-inverse. 

At the step ${t_0}$, the variational distribution is defined as:

\begin{equation}
	\label{eq:8}
	{p_{{{\bf{\theta }}_{{t_0}}}}}\left( {{\bf{\bar x}}_{{t_0}}^{(i)}|{\bf{y}}} \right) = \left\{ {\begin{array}{*{20}{l}}
			{{\cal N}({{{\bf{\bar y}}}^{(i)}},\sigma _{{t_0}}^2{\bf{I}}})&{{\rm{if }}\ {s_i} > 0}\\
			{{\cal N}(0,\sigma _{{t_0}}^2{\bf{I}}})&{{\rm{if }}\ {s_i} = 0}
	\end{array}} \right.
\end{equation}

At the step $t < {t_0}$, the variational distribution is defined as:

\begin{equation}
	\label{eq:9}
	p_{{\theta _t}}^{(t)}({\bf{\bar x}}_t^{(i)}|{{\bf{x}}_{t + 1}},{\bf{y}}) = \left\{ {\begin{array}{*{20}{l}}
			{{\cal N}\left( {{{{\bf{\bar y}}}^{(i)}},\sigma _t^2{\bf{I}}} \right)\ {\rm{ if }}\ {s_i} > 0}&{}\\
			{{\cal N}\left( {{\bf{\bar x}}_{{\theta _t}}^{(i)},\sigma _t^2{\bf{I}}} \right)\ {\rm{  if }}\ {s_i} = 0}&{}
	\end{array}} \right.
\end{equation}
where  ${\sigma _t}$ is the variance of the diffusion noise in ${{\bf{x}}_t}$. Once we obtain the values ${{\bf{x}}_{{\theta _t}}}$, we can multiply the matrix ${{\bf{V}}^T}$  with ${{\bf{x}}_{{\theta _t}}}$  and obtain  ${{\bf{\bar x}}_{{\theta _t}}}$. Then  ${\bf{\bar {\bf{x}}}}_t^{(i)}$ can be sampled from \eqref{eq:9}, and we can obtain the result ${{\bf{x}}_t}$  by multiplying $\bf{V}$  with  ${{\bf{\bar {\bf{V}}}}_t}$. Iteratively, the final result  ${{\bf{x}}_0}$ is generated. To summarize, the reverse diffusion process of S2DM is presented in Algorithm 1. The parameter updating of the network and the diffusion process are iteratively performed.

\begin{table}[t]
	\centering  
	\begin{tabular}{l} % four columns  
		\toprule[2pt]  %begin the first line  
		\textbf{Algorithm 1}: Reverse Diffusion Process for S2DM  \\  
		\hline %begin the second line  
		\textbf{Input}: The incomplete 3D seismic data ${\bf {y}}$, the hyperparameter $t_0$, $T$, $\beta_{1:T}$, $\sigma_{1:T}$\\
		\ \ 1. Initial the parameters of the network $\theta_{t_0}$\\
		\ \ 2. Generate the sample ${\bf {x}}_{t_0}$ by reparameterizing the distribution in equation\eqref{eq:8}\\
		\ \ 3. \textbf{For} $t=t_0-1, t_0-2,...,1$:\\
		\ \ 4.      \ \ \ \ \ \ Update parameters $\theta_{t}$ via equation \eqref{eq:6}\\
		\ \ 5. \ \ \ \ \ \ Predict ${\bf {x}}_{\theta_{t}}$ using the network with the parameters $\theta_{t}$\\
		\ \ 6. \ \ \ \ \ \ Obtain ${\bf {x}}_{{t-1}}$ by reparameterizing the distribution in equation\eqref{eq:9}\\
		\ \ 7. \textbf{End}\\
		\textbf{Output}: The reconstructed sample ${\bf {x}}_{0}$  \\  \bottomrule[2pt] %begin the third line 
	\end{tabular}  
\end{table}

\section*{Results}
\subsection{Experimental Settings}
To verify the performance of our method, we compare S2DM with three benchmark algorithms: Damped Rank Reduction (DRR), Deep Prior-Based MultiResUNet (DIP), and 3-DPCNN. DRR is a traditional method. DIP and 3-DPCNN are deep learning-based methods, employing unsupervised or self-supervised learning. The hyperparameters of these comparison methods are set according to their original paper \cite{17}, \cite{26}, and \cite{28}, respectively.

We evaluate the performance using four datasets. The first one is the SEG C3, a synthetic dataset with a 0.008s sampling interval. The second and third datasets named Parihaka-3D and Kerry-3D, are poststack field data with sampling intervals of 0.003 s and 0.004 s, respectively, provided by New Zealand Crown Minerals. The last dataset is Mobil Avo Viking Graben Line 12, which is prestack field data with a sampling interval of 0.004s. For all experiments, the size of the data blocks used is $128 \times 32 \times 128$ (inline $\times$ xline $\times$ time). All datasets are sourced from \href{}{https://wiki.seg.org/wiki/Open\_data.}

To evaluate the quality of reconstruction results, we quantitatively evaluate all experiments in this paper using Peak Signal-to-Noise Ratio (PSNR) and Structural Similarity (SSIM) metrics. PSNR is one of the most common methods for the assessment of the reconstruction quality and is calculated by the following formula:

 \begin{equation}
	\label{eq:10}
{\rm{PSNR}} = 10{\lg}\left( {\frac{{{{\max }^2}}}{{{\rm{MSE}}}}} \right)
\end{equation}
where $\max$ represents the maximum value of the reconstructed data. Mean squared error (MSE) is given by:

 \begin{equation}
	\label{eq:11}
{\rm{MSE}}=\frac{1}{M}\sum\limits_{m=1}^{M}{{{\left( {\bf {x}}_m-{{\bf {r}}_m}\right)}^{2}}}
\end{equation}
Where ${\bf {r}}_m$ and ${\bf {x}}_m$ denote the $m$th pixel in the reconstructed result ${\bf {r}}$ and complete data ${\bf {x}}$, respectively. $M$ is the total number of pixels in the sample ${\bf {x}}$.

SSIM is used to evaluate the spatial similarity between two samples. The expression is as follows:

 \begin{equation}
	\label{eq:12}
{\rm{SSIM}} = \frac{{\left( {2{\mu _r}{\mu _x} + {c_1}} \right)\left( {2{\sigma _{rx}} + {c_2}} \right)}}{{\left( {\mu _r^2 + \mu _x^2 + {c_1}} \right)\left( {\sigma _r^2 + \sigma _x^2 + {c_2}} \right)}}
\end{equation}
Where ${\mu _{r}}$ and ${\mu _{x}}$ are the means of the reconstructed result ${\bf {r}}$ and the incomplete data ${\bf {x}}$, ${\sigma ^2_{r}}$ and ${\sigma ^2_{x}}$ are their variances and ${\sigma _{rx}}$ is the covariance between them. Constants ${c _{1}}$ and ${c _{2}}$ are introduced to stabilize the fraction. An SSIM value closer to 1 indicates a higher similarity to the original data. We use the functions provided by the scikit-image library to calculate PSNR and SSIM, with default parameters setting for ease of comparison.

In these experiments, we set $T$ to 50,000 and the starting time ${t_0}$ for the reverse process to 100. For the hyperparameter ${\beta _{1:T}}$ representing the variance of Gaussian noise, we linearly generate 50,000 values between ${10 ^{-5}}$ and $5 \times {10 ^{-5}}$. Additionally, the network iterates twice for each step in the reverse diffusion process. Our network utilizes Adam as the optimizer and initializes the learning rate of ${10 ^{-4}}$. These settings are applied consistently to all.

\subsection{Results and Analysis}

Figure 4-5 presents the reconstruction results of SEG C3. At a 50\% missing rate as shown in Figure 4, the result reconstructed by DRR (Figure 4b) shows low consistency, and there are some obvious remaining signals in the residual image (Figure 4g). Contrastively, three deep learning-based methods could reconstruct the signals integrally. As shown in the three residual images for the three deep-learning methods (Figure 4h-4j), we just observe a small amount of signal leakage, such as the area in the green rectangle. In comparison, our S2DM shows the greatest reconstruction ability among the three deep-learning methods, that is proven by the lowest amplitude of signal leakage shown in Figure 4j. Figure 5 presents the reconstructed results shown in 2D slices. We can clearly observe the performance advantages of S2DM. Especially in the area within the green rectangle, it is easy to see the signal leakage to varying degrees in the compared methods. However, there is almost little leakage in S2DM.

Figure 6 gives the reconstructed results in a single trace version at a 50\% missing rate. 
All methods could basically reconstruct the missing trace both in amplitude and phase. Additionally, the reconstructed waveform of S2DM is more similar to the true waveform compared with the three competitive methods, such as the peaks from 100 to 120. Thus, S2DM performs better than the compared methods in terms of a single trace. 

Table 1 displays PSNR and SSIM metrics of the reconstructed results. It is evident that the proposed S2DM achieves the best results in these evaluation metrics. Benefiting from the diffusion process, the proposed S2DM could generate many complex samples based on the input data to train the network, thereby enhancing the performance of VTSM. Meanwhile, with the aid of VTSM, missing traces can be reconstructed gradually by sampling the posterior distribution of data given the input measurements.

To further assess the performance and verify the performance of our method, we test on the field dataset, Parihaka-3D. Similar to the synthetic data, we set the missing rate at 50\%. The reconstruction results are presented in Figures 7-8. All methods could reconstruct missing traces to some extent as shown in Figure 7b-7e and Figure 8b-8e. However, we can observe that there are fewer remaining effective signals in the residual image of the S2DM than those of DRR, DIP, and 3-DPCNN, such as the region in the green rectangle in Figure 7g-7f and Figure 8g-8f.

Figure 9 compares the reconstruction performance of DRR, DIP, 3-DPCNN and S2DM under the same missing trace. We can observe that the reconstructed trace of the proposed S2DM can match the ground truth more accurately, especially in the peaks and troughs.

Additionally, Table 2 presents PSNR and SSIM metrics of the reconstructed data. Clearly, the S2DM method has the highest evaluation metrics across different missing rates. Generally speaking, the proposed S2DM shows superior reconstruction performance to the compared methods in the poststack data Parihaka-3D.

The experiments above validate S2DM's reconstruction capability under random missing. In the following, we attempt to reconstruct more types of seismic datasets to verify the model's generalization ability.

Firstly, we fold the prestack dataset, Mobil Avo Viking Graben Line 12, into 3D for experimentation. As shown in Figure 10 and Figure 11, all methods could reconstruct the missing traces in the prestack field data to some extent. However, we can still observe that there are fewer effective signals in the residual image of S2DM compared with those in the compared methods. Table 3 presents the PSNR and SSIM metrics, further verifying the superiority of the proposed S2DM over other methods in the prestack data.

Then, we conduct some experiments using Kerry-3D dataset. The Kerry-3D dataset contains numerous faults, causing a significant challenge for the reconstruction task. Figure 12 shows that S2DM provides clearer and more complete reconstruction results compared with other methods. The residual images reflect the superior reconstruction performance of S2DM. The effectiveness of S2DM is more obvious in the 2D slices, as illustrated in Figure 13.

\iffalse
\begin{table}[!htbp]  \caption{\textbf{PSNR and SSIM on 1) Mobil Avo Viking Graben Line 12 and 2) Kerry-3D.}}%title  
	\centering  
	\begin{tabular}{ccccc} % four columns  
		\toprule[2pt]  %begin the first line  
		{}& PSNR  1)& SSIM 1)&PSNR  2)&SSIM 2)\\  
		\hline %begin the second line  
		DRR&35.1610dB&0.9745&32.1710dB&0.9462\\
		DIP &36.5702dB&0.9810&37.1286dB&0.9775\\
		3-DPCNN&37.1464dB&0.9844&35.9801dB&0.9807\\
		S2DM&37.4179dB&0.9857&38.6564dB&0.9851\\ \bottomrule[2pt] %begin the third line 
	\end{tabular}  
\end{table}
\fi

\section*{Discussion}
\subsection{Performance Analysis at Different Missing Rates}
To analyze the performance of the proposed S2DM at different random missing rates, we perform some experiments on Parihaka-3D, and the results are provided in Table 3. From Table 3, it is evident that the abilities of all reconstruction methods degrade with the missing rate raises. However, we can observe that the proposed S2DM exhibits a superior reconstruction performance to the compared methods from the missing rate of 50\% to 90\%. Even at the high missing level (90\%), the proposed S2DM still achieves the best performance among all methods. Generally speaking, the proposed S2DM achieves a good reconstruction performance at different missing levels.

\subsection{Time Cost Analysis}
To evaluate the time cost of S2DM, we record the computing time of different methods to complete the reconstruction task on Parihaka-3D dataset. All methods are run in the same environment with an Intel Xeon Gold 5320 CPU (2.2GHz) and an NVIDIA A30 GPU (24GB). The collected results are shown in Table 4. It is evident that DRR and DIP need less time than 3-DPCNN and S2DM. 3-DPCNN consumes about six times the time of S2DM. DRR is a traditional optimization method without the training procedure, thereby consuming relatively little time. S2DM trains the network $t_0$ times repeatedly, whereas DIP only trains the network once. Thus, S2DM consumes more time than DIP. Considering that the reconstruction performance of S2DM is much better than DIP, it is acceptable for such a relatively high time cost. Moreover, S2DM is superior to the latest self-supervised reconstruction method 3-DPCNN in terms of time cost.

\subsection{Advantages and Disadvantages}
In general, the proposed S2DM based on the diffusion model has three significant advantages: firstly, the proposed S2DM operates in a self-supervised learning mechanism, without demand for extra supervised information or training data, reducing the time and economic cost. Secondly, the proposed S2DM combines the diffusion process with the 3D network, enhancing the performance of S2DM in dealing with seismic data with complex distributions. The proposed S2DM generates good reconstruction results in different seismic data and at different missing rates. Thirdly, the proposed S2DM only consumes nearly one-sixth of the time but exhibits an obvious performance benefit relative to the latest 3-DPCNN. 

Except for the above advantages, the proposed S2DM also appears some disadvantages. Firstly, the proposed S2DM has some difficulties in dealing with seismic data with continuous large missing traces as the receptive field of the convolution kernel used in S2DM is finite, making it hard to cover missing regions. Secondly, although the proposed S2DM consumes less time than 3-DPCNN, it needs more time than DIP. In the future, we will focus on these disadvantages to further perfect the reconstruction method.

\section*{Conclusion}
In this article, we present S2DM as a method for 3D seismic data reconstruction. By virtue of a diffusion restoration model and a variational time-spatial module, the proposed S2DM could reconstruct randomly missing traces integrally with no extra labeled-training data. Specifically speaking, the proposed S2DM introduces a variational-interference-based loss function as the deep-prior to optimize the untrained 3D time-spatial network in a self-supervised manner without any need for extra labeled data. Additionally, the proposed S2DM gradually complicates the distribution of input data in the forward diffusion process, enhancing the generalization performance of the network. Furthermore, with the aid of VTSM, missing traces could be reconstructed gradually in the reverse diffusion process. Results on one synthetic dataset and three representative field datasets demonstrate that the proposed S2DM could reconstruct missing traces with higher PSNR, SSIM and less signal leakage compared with well-known DRR, DIP, and 3-DPCNN methods.

\bibliographystyle{seg.bst}
\bibliography{example}

\newpage
\section{List of Figures}

Figure 1: The structure of the variational time-spatial network.\\
Figure 2: The structure of the variational time-spatial network.\\
Figure 3: The structures of AB and RB block.\\
Figure 4: Reconstruction results of the synthetic data. (a) Ground truth of the SEG C3. (b)-(e) Reconstruction results for DRR, DIP, 3-DPCNN, and S2DM, respectively. (f) Ground truth at 50\% randomly missing traces. (g)-(j) Residuals corresponding to DRR, DIP, 3-DPCNN, and S2DM.\\
Figure 5: (a)-(j) 2D slices of the data shown in Figure 4a-4j, respectively.\\
Figure 6: Trace comparisons at a missing position for DRR, DIP, 3-DPCNN and S2DM, respectively.\\
Figure 7: Reconstruction results of the poststack field data. (a) Ground truth of Parihaka-3D. (b)-(e) Reconstruction results for DRR, DIP, 3-DPCNN, and S2DM, respectively. (f) Ground truth with 50\% randomly missing traces. (g)-(j) Residuals corresponding to DRR, DIP, 3-DPCNN, and S2DM.\\
Figure 8: (a)-(j) 2D slices of the data shown in Figure 7a-7j, respectively.\\
Figure 9: Trace comparisons at a missing position for DRR, DIP, 3-DPCNN and S2DM, respectively.\\
Figure 10: Reconstruction results of the prestack field data. (a) Ground truth from Mobil Avo Viking Graben Line 12. (b)-(e) Reconstruction results for DRR, DIP, 3-DPCNN, and S2DM, respectively. (f) Ground truth with 50\% randomly missing traces. (g)-(j) Residuals corresponding to DRR, DIP, 3-DPCNN, and S2DM.\\
Figure 11: (a)-(j) 2D slices of the data displayed in Figure 10a-10j, respectively.mparisons at a missing position for DRR, DIP, 3-DPCNN and S2DM, respectively.\\
Figure 12: Reconstruction results of the poststack field data. (a) Ground truth from Kerry-3D. (b)-(e) Reconstruction results for DRR, DIP, 3-DPCNN, and S2DM, respectively. (f) Ground truth with 50\% randomly missing traces. (g)-(j) Residuals corresponding to DRR, DIP, 3-DPCNN, and S2DM.\\
Figure 13: (a)-(j) 2D slices of the data displayed in Figure 12a-12j, respectively.\\
\section{List of Tables}
Table 1: PSNR and SSIM on 1) SEG C3 and 2) Parihaka-3D.\\
Table 2: PSNR and SSIM on 1) Mobil Avo Viking Graben Line 12 and 2) Kerry-3D.\\
Table 3: PSNR and SSIM on Parihaka-3D at different missing levels.\\
Table 4: Computing time comparison.

\begin{figure}[htpb]
	\centering
	{\includegraphics[width=0.99\linewidth]{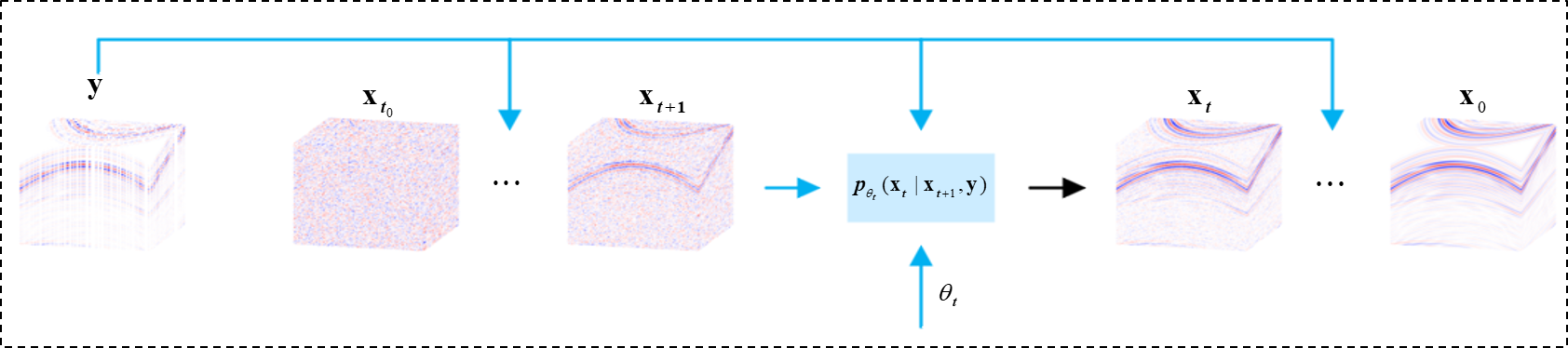}}
	\caption{An overview of S2DM.}
\end{figure}
\begin{figure}[htpb]
	\centering
	{\includegraphics[width=0.99\linewidth]{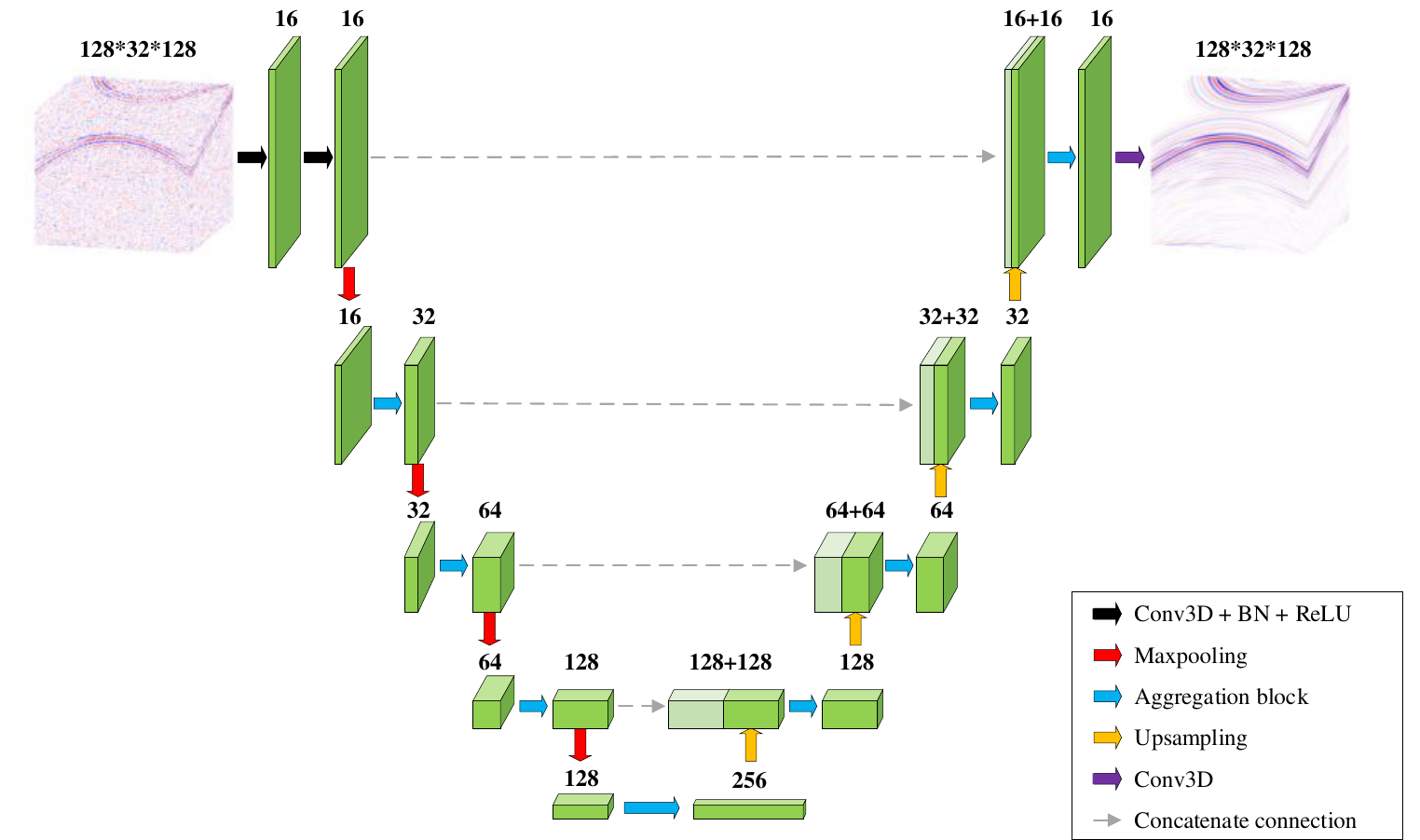}}
	
	\caption{The structure of the variational time-spatial network.}
\end{figure}
\begin{figure}[htpb]
	\centering
	{\includegraphics[width=0.99\linewidth]{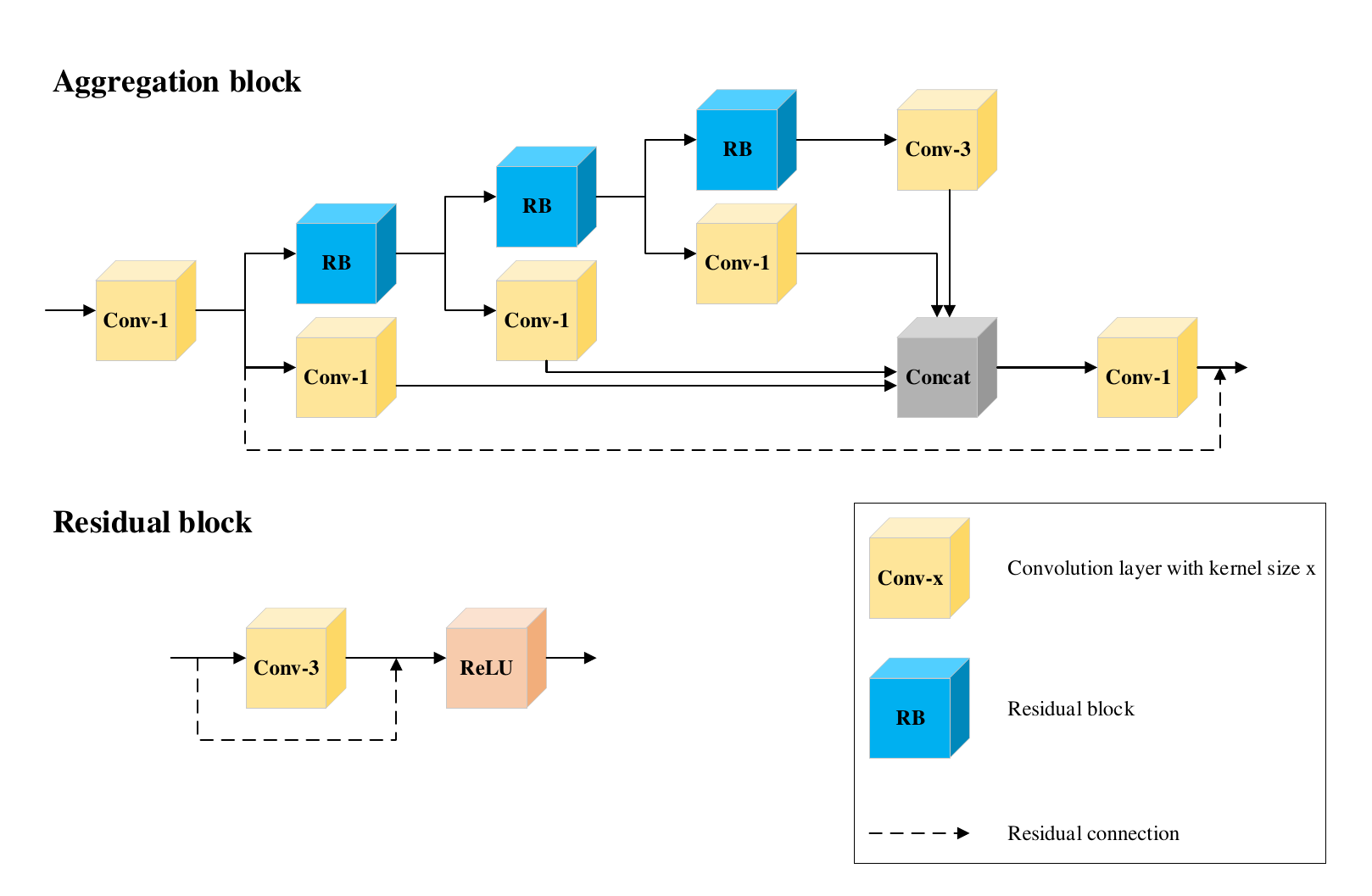}}
	
	\caption{The structures of AB and RB block.}
\end{figure}
\begin{figure}[htpb]
	\centering
	\subfloat[]{\includegraphics[width=0.2\linewidth]{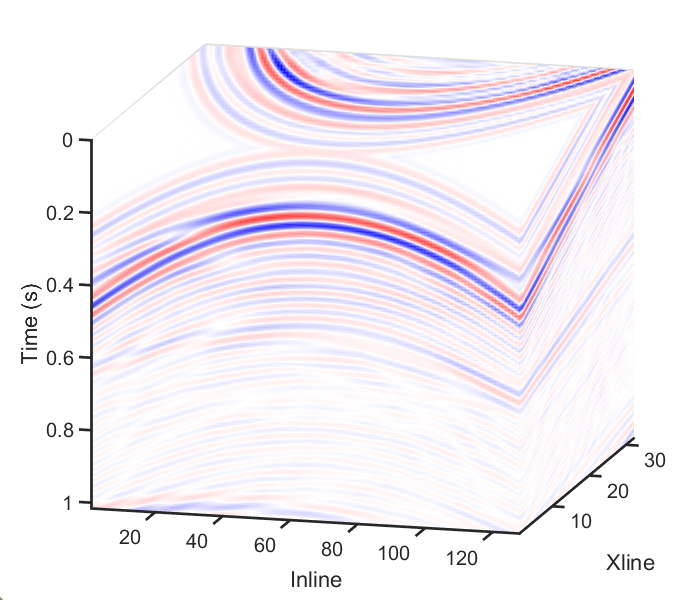}}
	\subfloat[]{\includegraphics[width=0.2\linewidth]{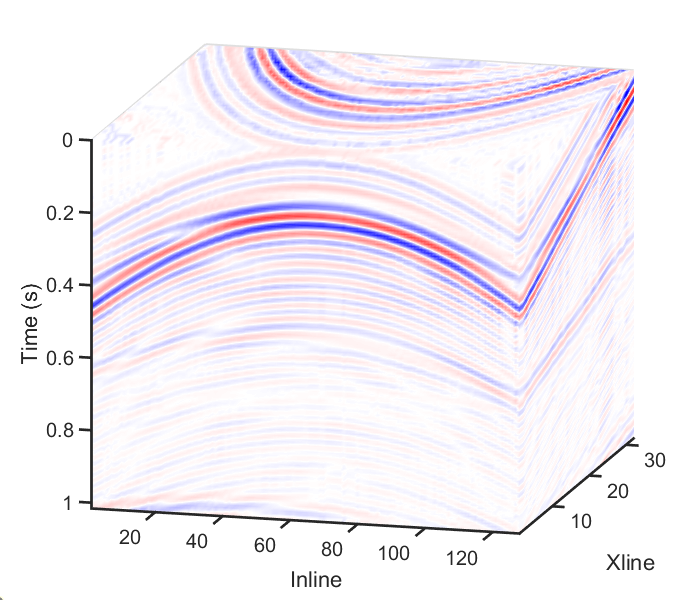}}
	\subfloat[]{\includegraphics[width=0.2\linewidth]{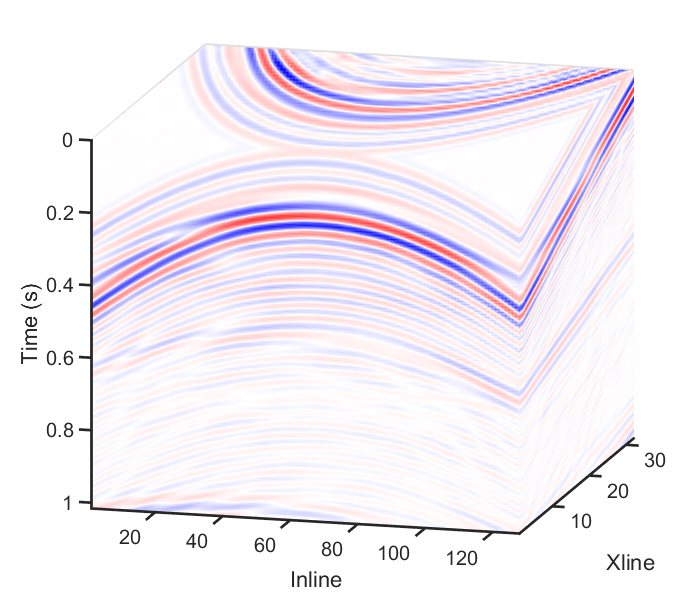}}
	\subfloat[]{\includegraphics[width=0.2\linewidth]{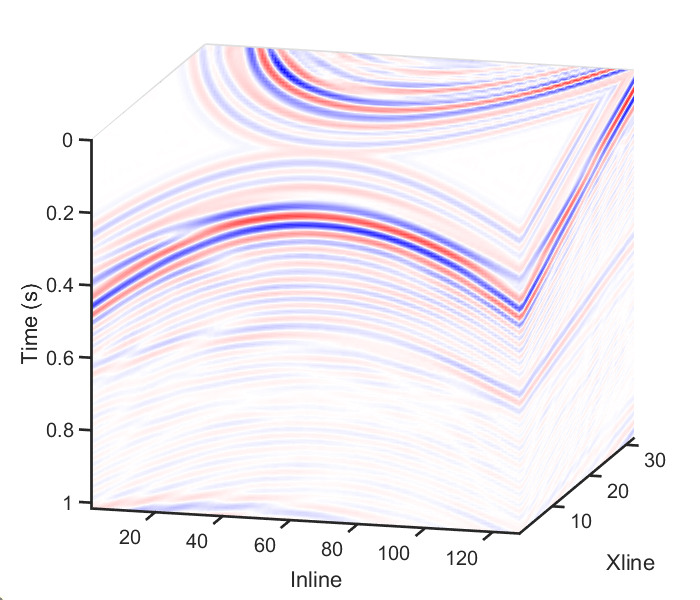}}
	\subfloat[]{\includegraphics[width=0.2\linewidth]{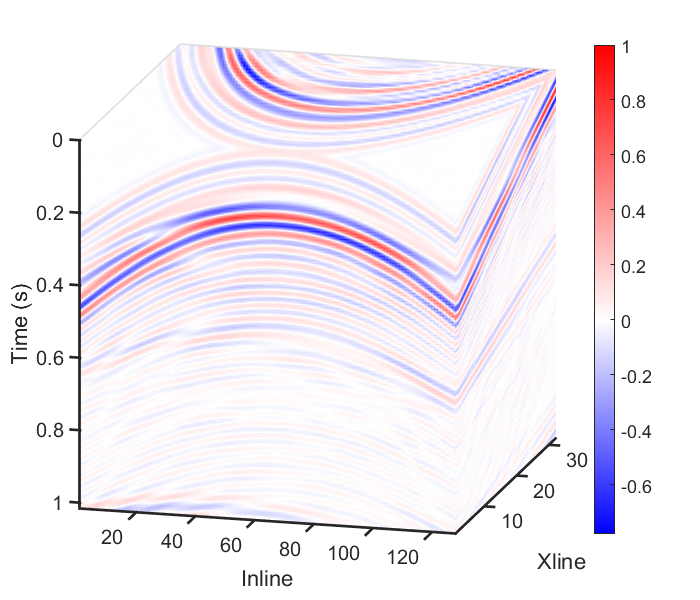}}
	\qquad
	\subfloat[]{\includegraphics[width=0.2\linewidth]{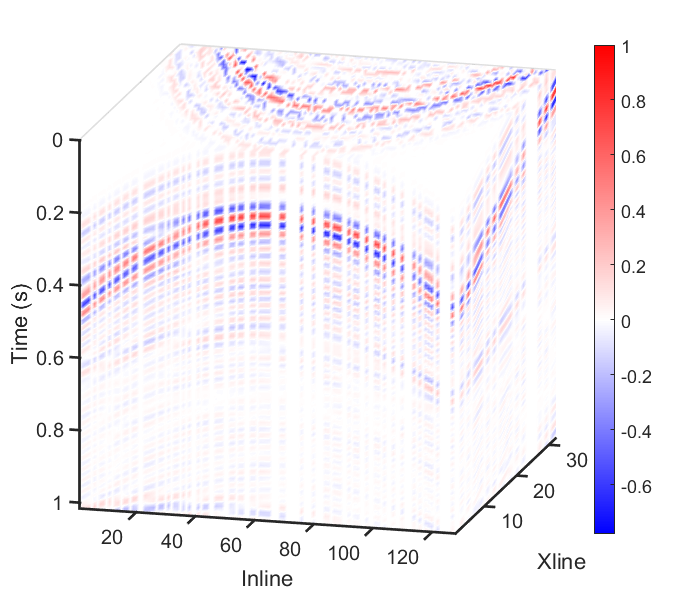}}
	\subfloat[]{\includegraphics[width=0.2\linewidth]{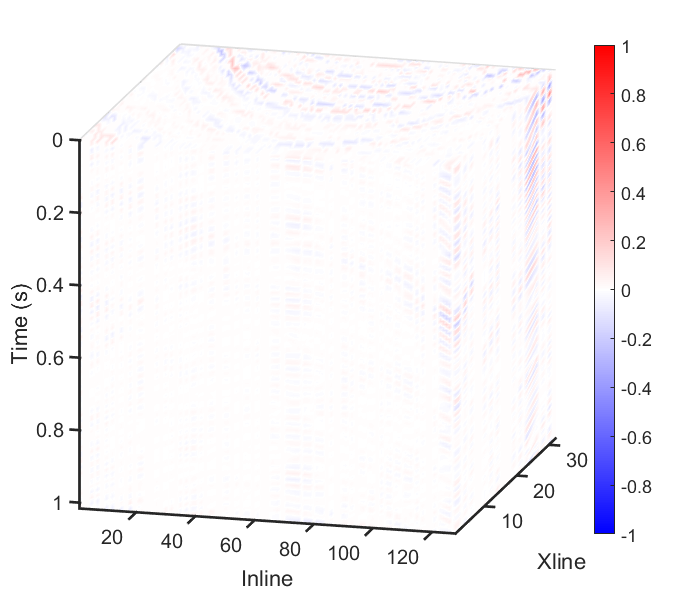}}
	\subfloat[]{\includegraphics[width=0.2\linewidth]{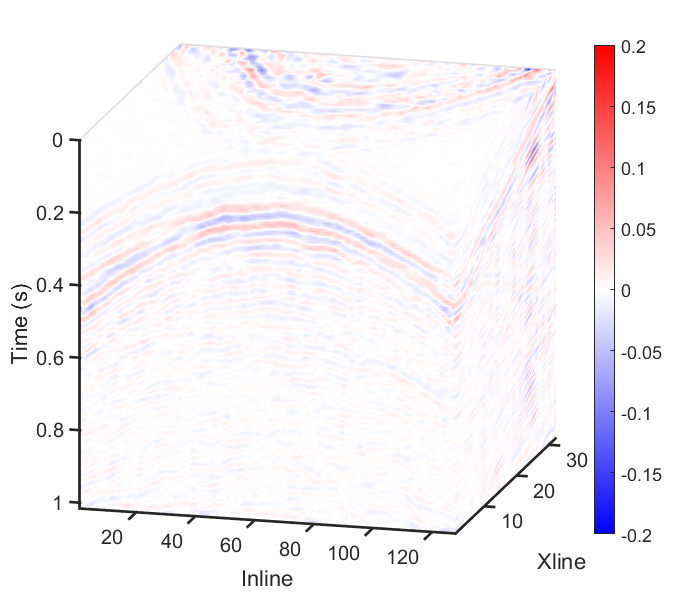}}
	\subfloat[]{\includegraphics[width=0.2\linewidth]{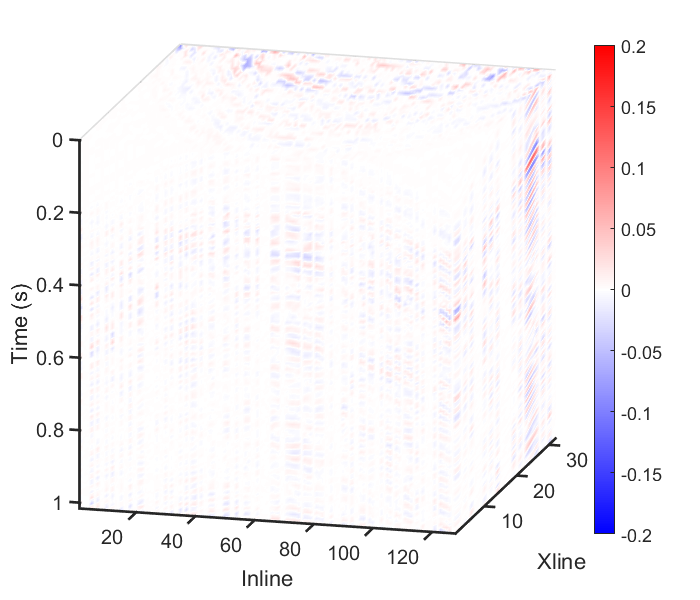}}
	\subfloat[]{\includegraphics[width=0.2\linewidth]{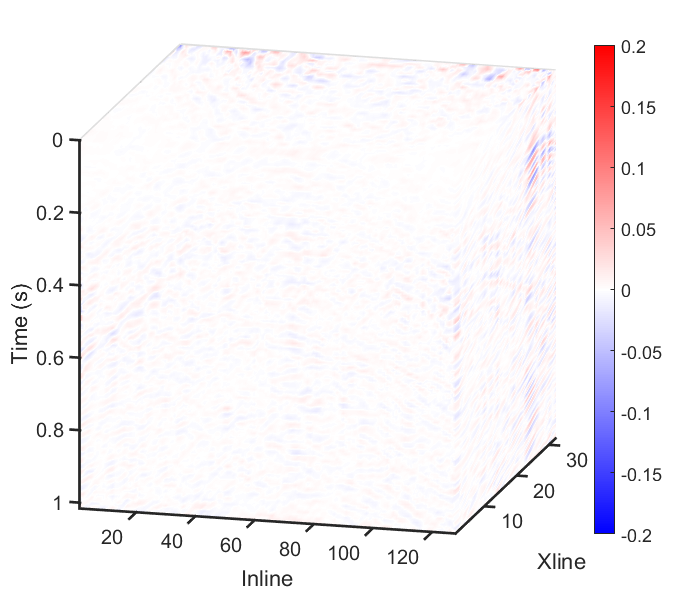}}
	\caption{Reconstruction results of the synthetic data. (a) Ground truth of the SEG C3. (b)-(e) Reconstruction results for DRR, DIP, 3-DPCNN, and S2DM, respectively. (f) Ground truth at 50\% randomly missing traces. (g)-(j) Residuals corresponding to DRR, DIP, 3-DPCNN, and S2DM.}
\end{figure}
\begin{figure}[htpb]
	\centering
	\subfloat[]{\includegraphics[width=0.2\linewidth]{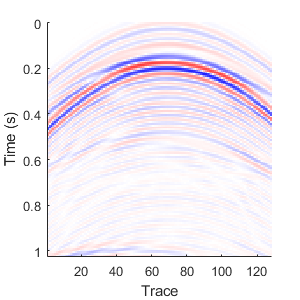}}
	\subfloat[]{\includegraphics[width=0.2\linewidth]{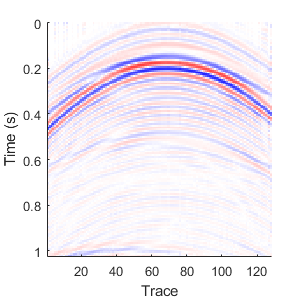}}
	\subfloat[]{\includegraphics[width=0.2\linewidth]{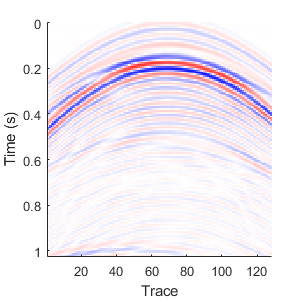}}
	\subfloat[]{\includegraphics[width=0.2\linewidth]{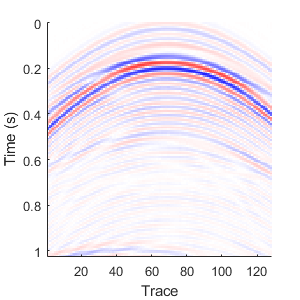}}
	\subfloat[]{\includegraphics[width=0.2\linewidth]{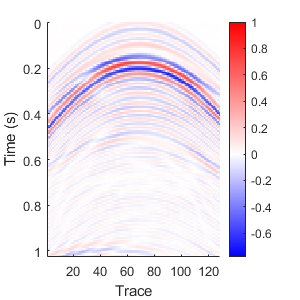}}
	\qquad
	\subfloat[]{\includegraphics[width=0.2\linewidth]{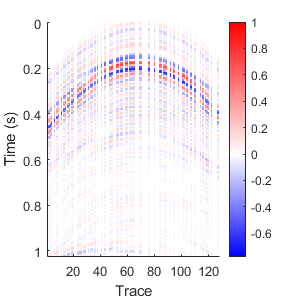}}
	\subfloat[]{\includegraphics[width=0.2\linewidth]{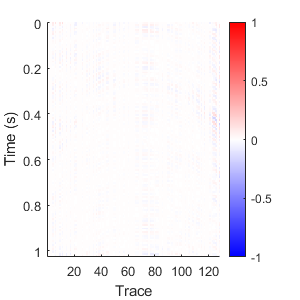}}
	\subfloat[]{\includegraphics[width=0.2\linewidth]{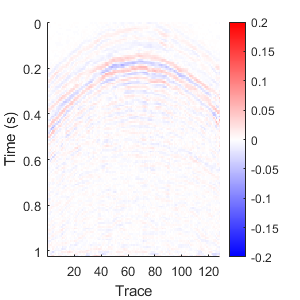}}
	\subfloat[]{\includegraphics[width=0.2\linewidth]{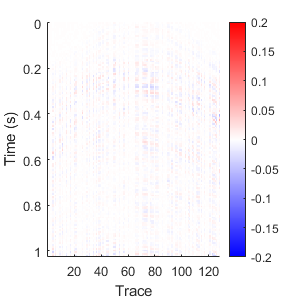}}
	\subfloat[]{\includegraphics[width=0.2\linewidth]{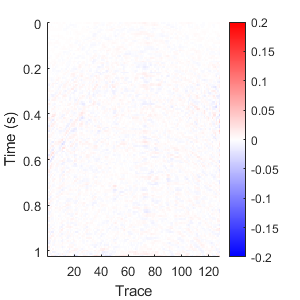}}
	
	\caption{(a)-(j) 2D slices of the data shown in Figure 4a-4j, respectively.}
\end{figure}
\begin{figure}[htpb]
	\centering
	{\includegraphics[width=0.99\linewidth]{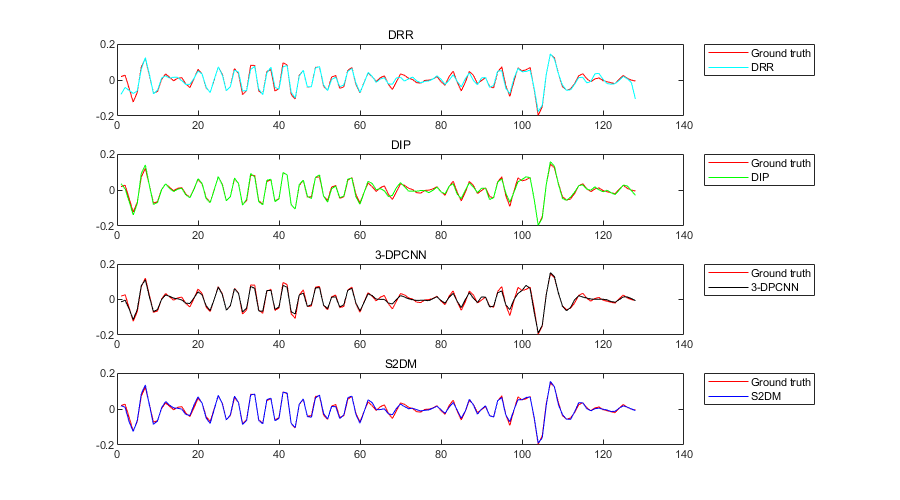}}
	
	\caption{Trace comparisons at a missing position for DRR, DIP, 3-DPCNN and S2DM, respectively.}
\end{figure}
\begin{figure}[htpb]
	\centering
	\subfloat[]{\includegraphics[width=0.2\linewidth]{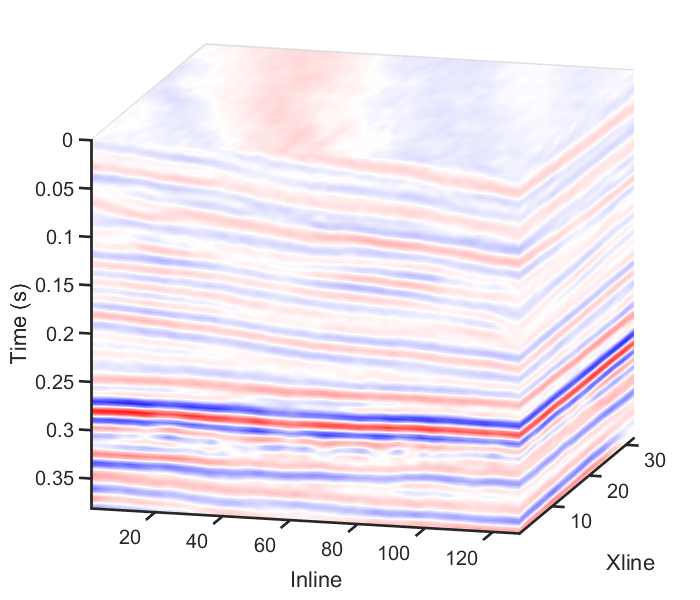}}
	\subfloat[]{\includegraphics[width=0.2\linewidth]{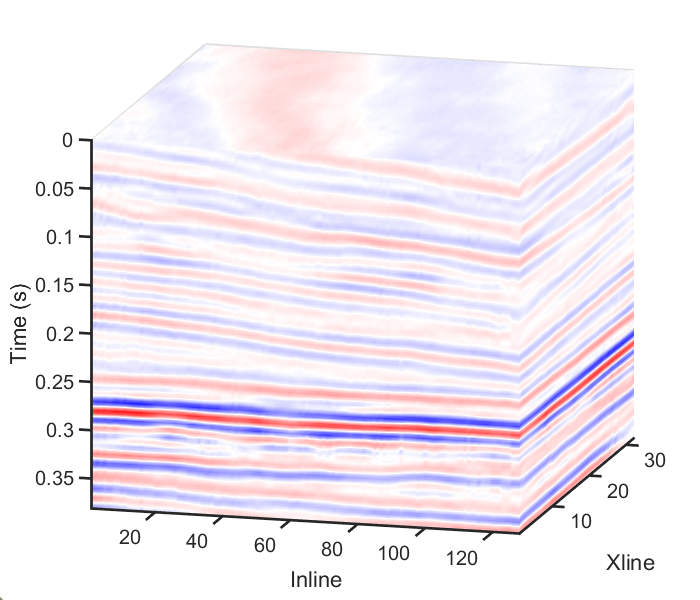}}
	\subfloat[]{\includegraphics[width=0.2\linewidth]{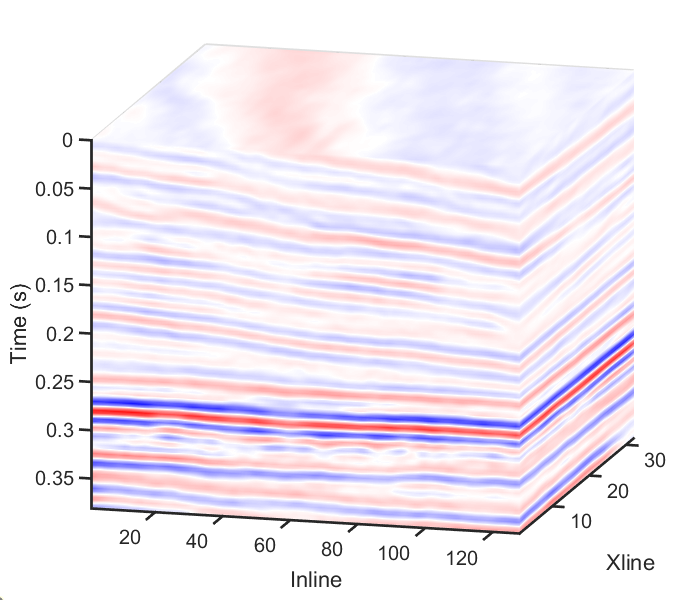}}
	\subfloat[]{\includegraphics[width=0.2\linewidth]{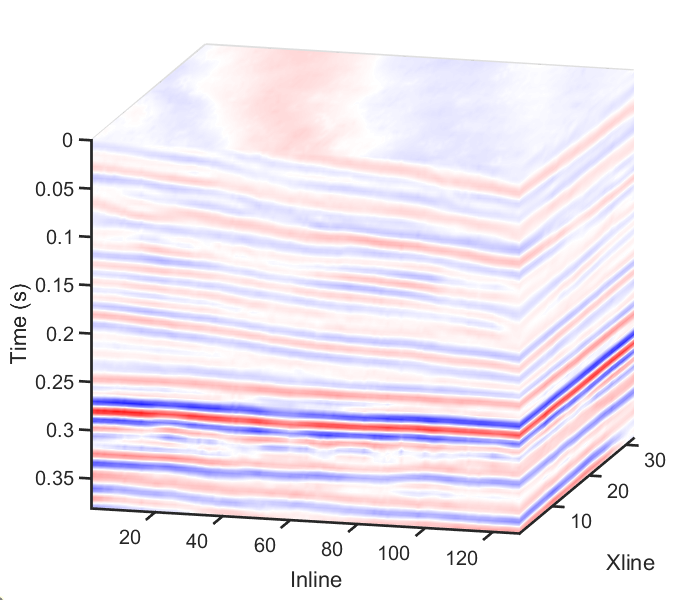}}
	\subfloat[]{\includegraphics[width=0.2\linewidth]{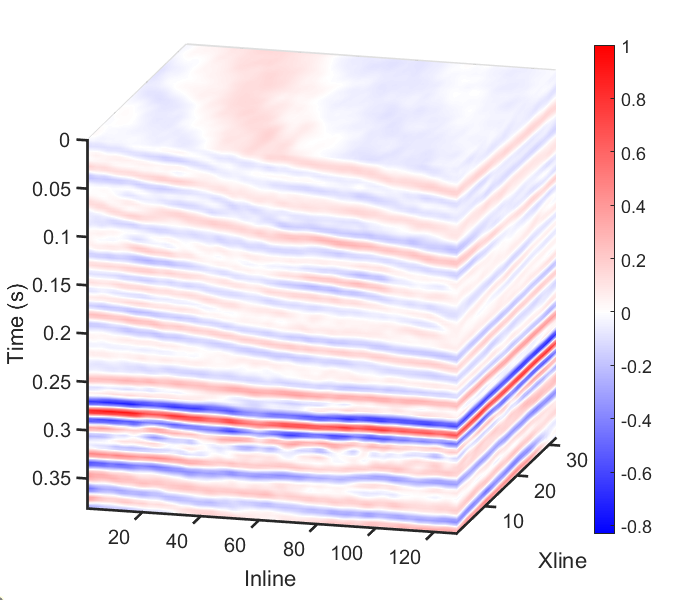}}
	\qquad
	\subfloat[]{\includegraphics[width=0.2\linewidth]{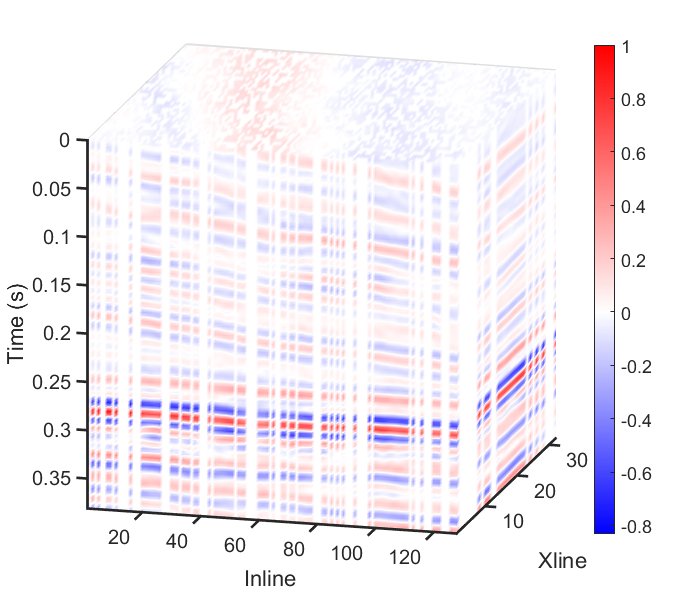}}
	\subfloat[]{\includegraphics[width=0.2\linewidth]{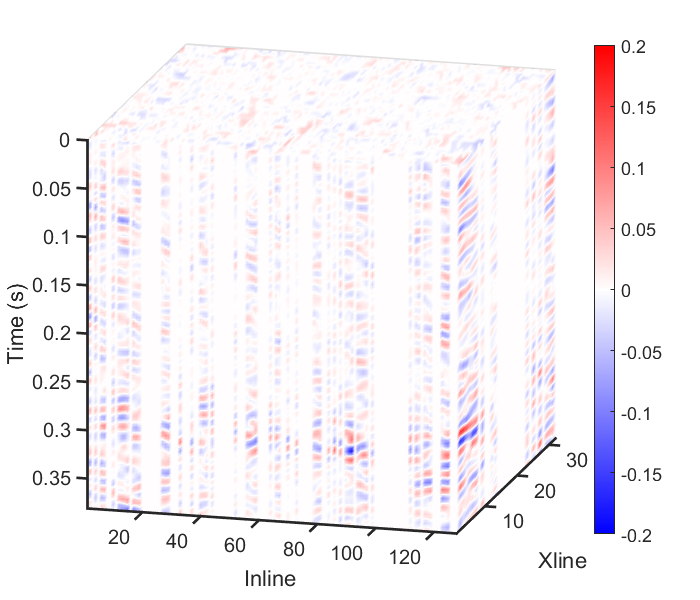}}
	\subfloat[]{\includegraphics[width=0.2\linewidth]{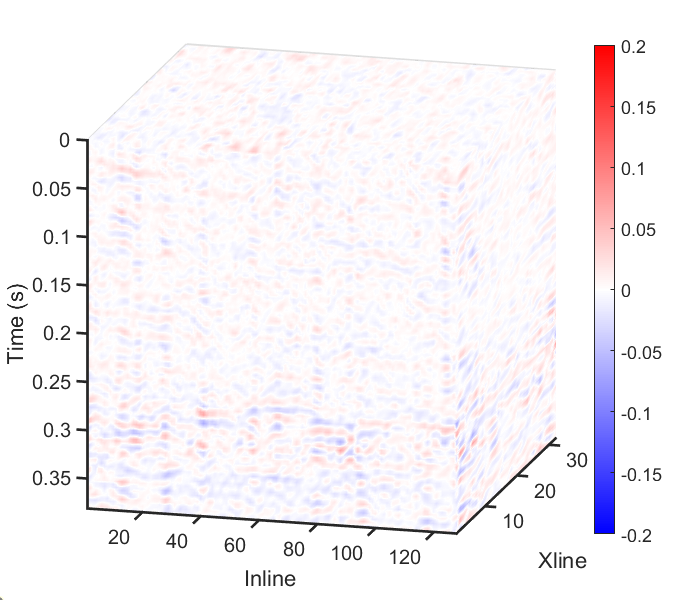}}
	\subfloat[]{\includegraphics[width=0.2\linewidth]{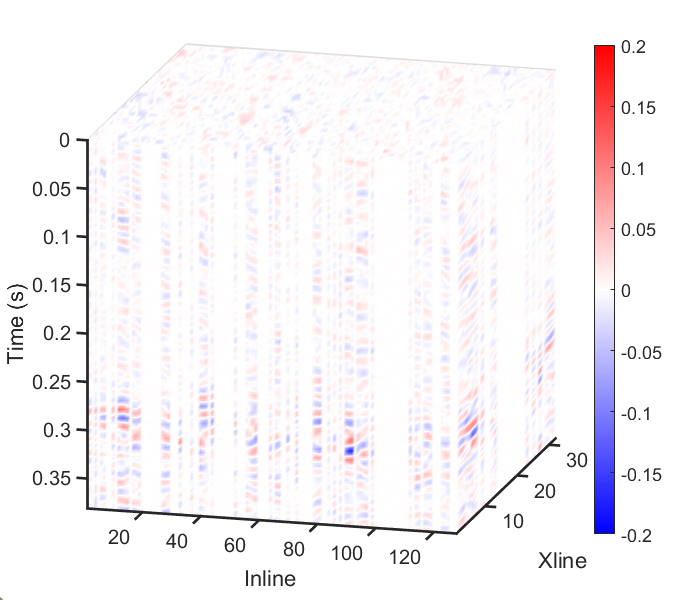}}
	\subfloat[]{\includegraphics[width=0.2\linewidth]{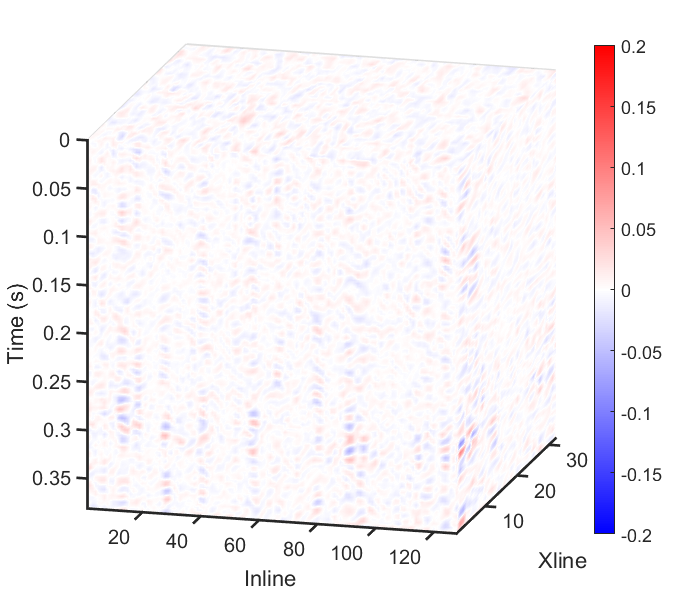}}
	\caption{Reconstruction results of the poststack field data. (a) Ground truth of Parihaka-3D. (b)-(e) Reconstruction results for DRR, DIP, 3-DPCNN, and S2DM, respectively. (f) Ground truth with 50\% randomly missing traces. (g)-(j) Residuals corresponding to DRR, DIP, 3-DPCNN, and S2DM.}
\end{figure}
\begin{figure}[htpb]
	\centering
	\subfloat[]{\includegraphics[width=0.2\linewidth]{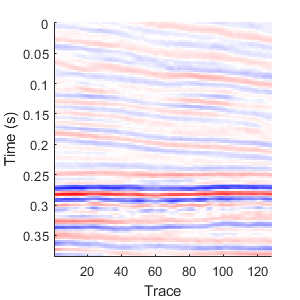}}
	\subfloat[]{\includegraphics[width=0.2\linewidth]{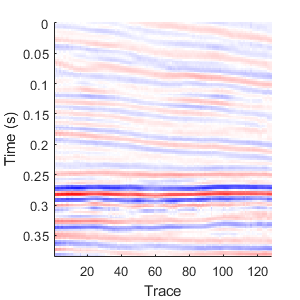}}
	\subfloat[]{\includegraphics[width=0.2\linewidth]{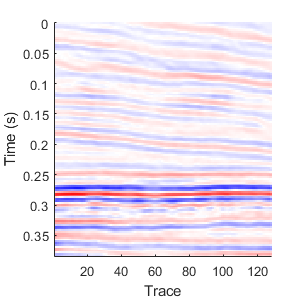}}
	\subfloat[]{\includegraphics[width=0.2\linewidth]{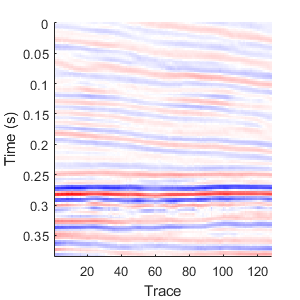}}
	\subfloat[]{\includegraphics[width=0.2\linewidth]{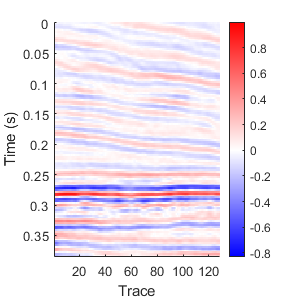}}
	\qquad
	\subfloat[]{\includegraphics[width=0.2\linewidth]{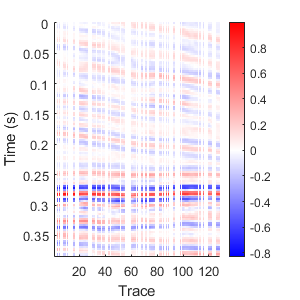}}
	\subfloat[]{\includegraphics[width=0.2\linewidth]{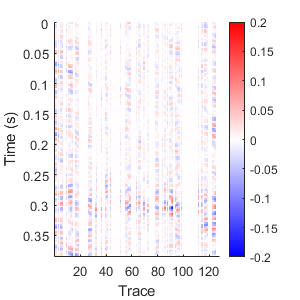}}
	\subfloat[]{\includegraphics[width=0.2\linewidth]{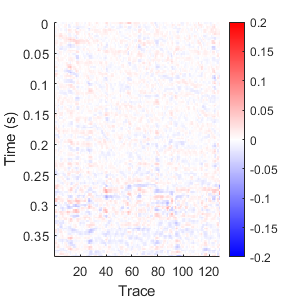}}
	\subfloat[]{\includegraphics[width=0.2\linewidth]{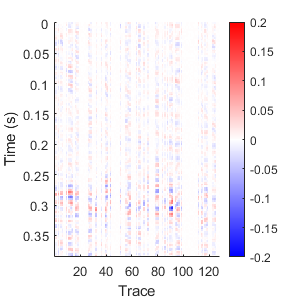}}
	\subfloat[]{\includegraphics[width=0.2\linewidth]{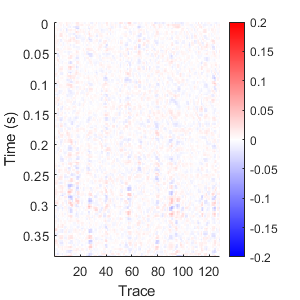}}
	
	\caption{(a)-(j) 2D slices of the data shown in Figure 7a-7j, respectively.}
\end{figure}
\begin{figure}[htpb]
	\centering
	{\includegraphics[width=0.99\linewidth]{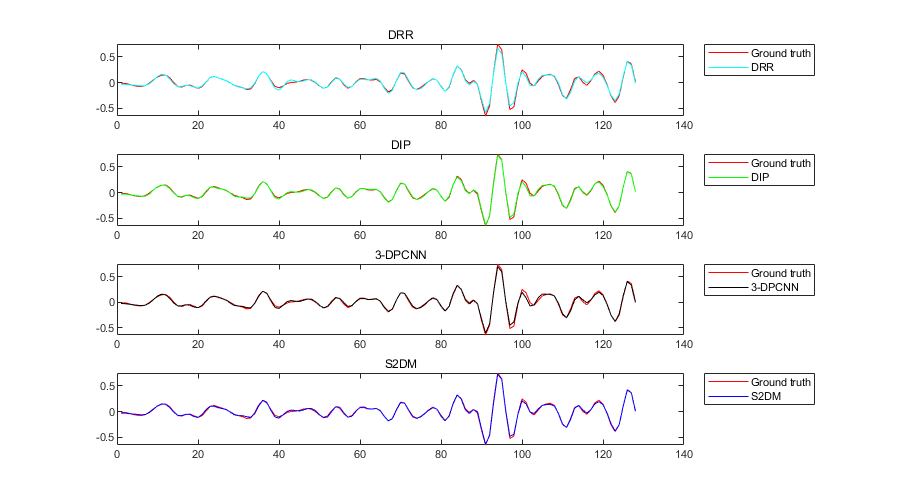}}
	\caption{Trace comparisons at a missing position for DRR, DIP, 3-DPCNN and S2DM, respectively.}
\end{figure}
\begin{figure}[htpb]
	\centering
	\subfloat[]{\includegraphics[width=0.2\linewidth]{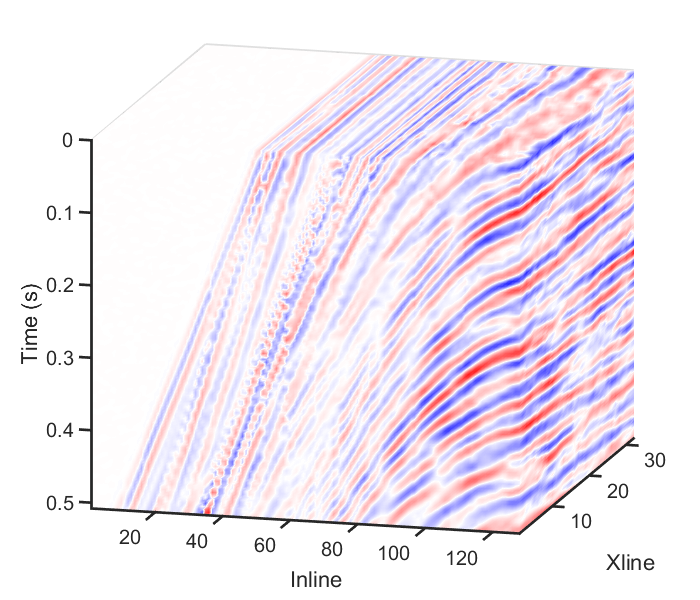}}
	\subfloat[]{\includegraphics[width=0.2\linewidth]{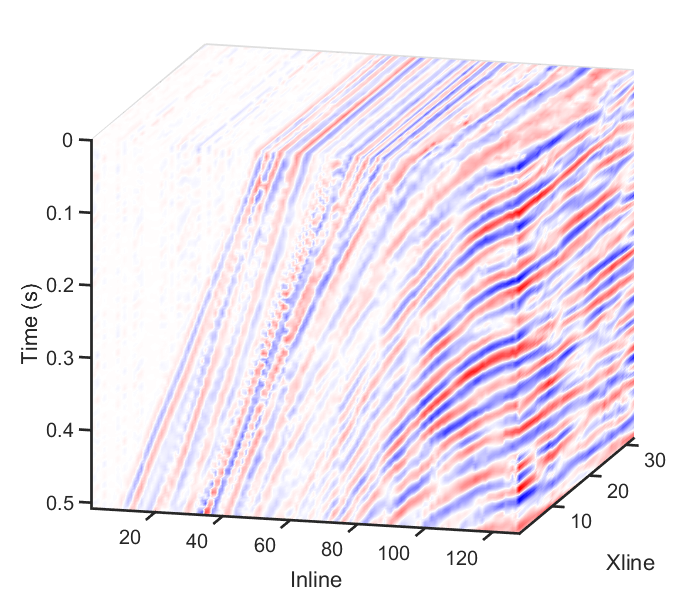}}
	\subfloat[]{\includegraphics[width=0.2\linewidth]{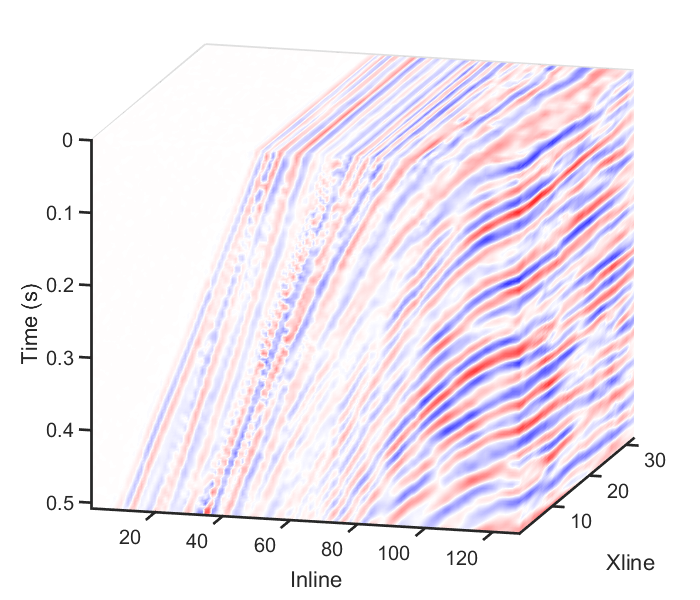}}
	\subfloat[]{\includegraphics[width=0.2\linewidth]{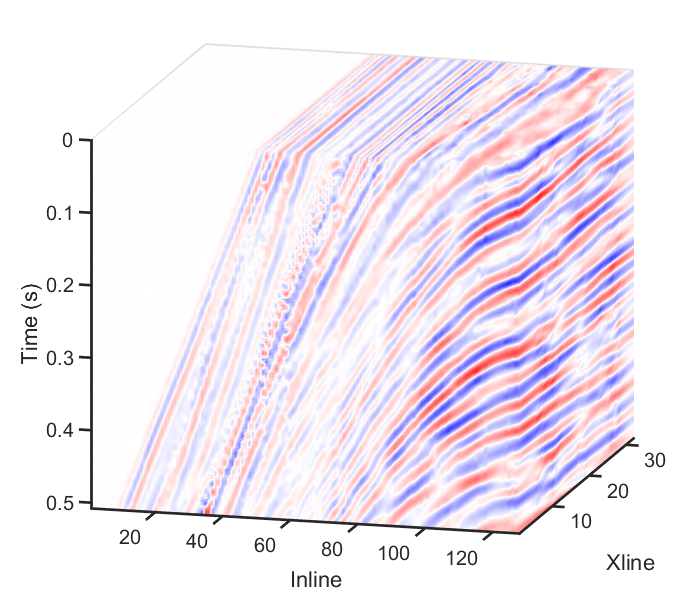}}
	\subfloat[]{\includegraphics[width=0.2\linewidth]{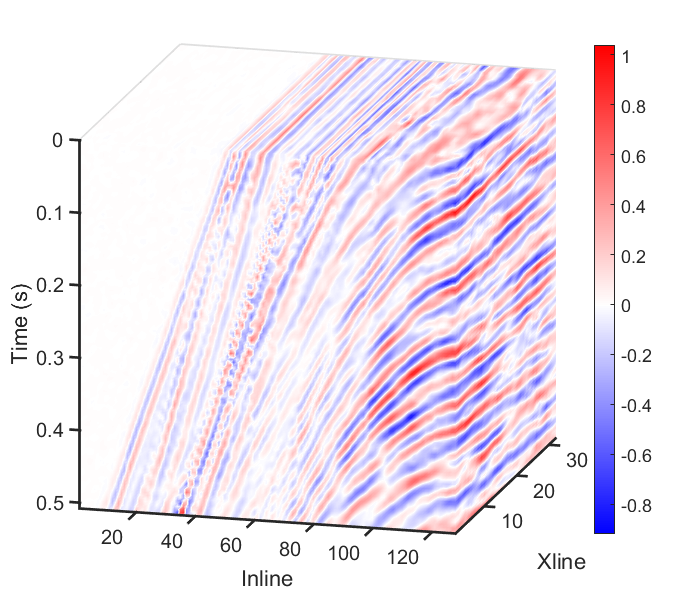}}
	\qquad
	\subfloat[]{\includegraphics[width=0.2\linewidth]{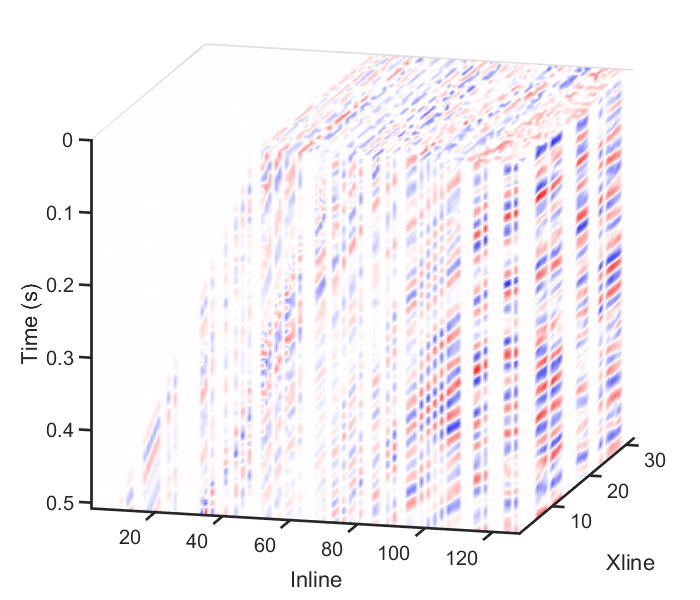}}
	\subfloat[]{\includegraphics[width=0.2\linewidth]{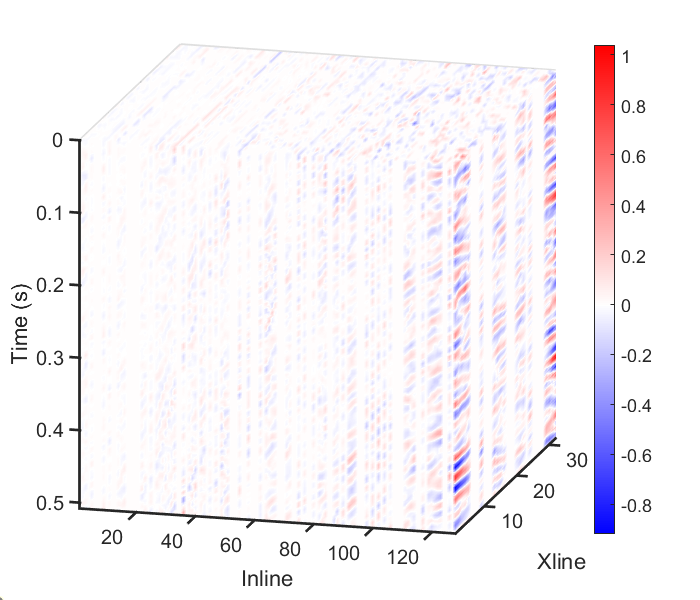}}
	\subfloat[]{\includegraphics[width=0.2\linewidth]{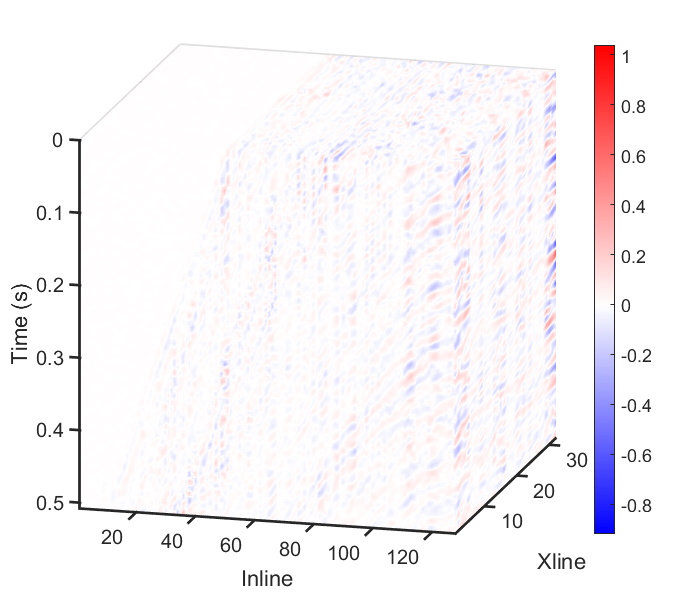}}
	\subfloat[]{\includegraphics[width=0.2\linewidth]{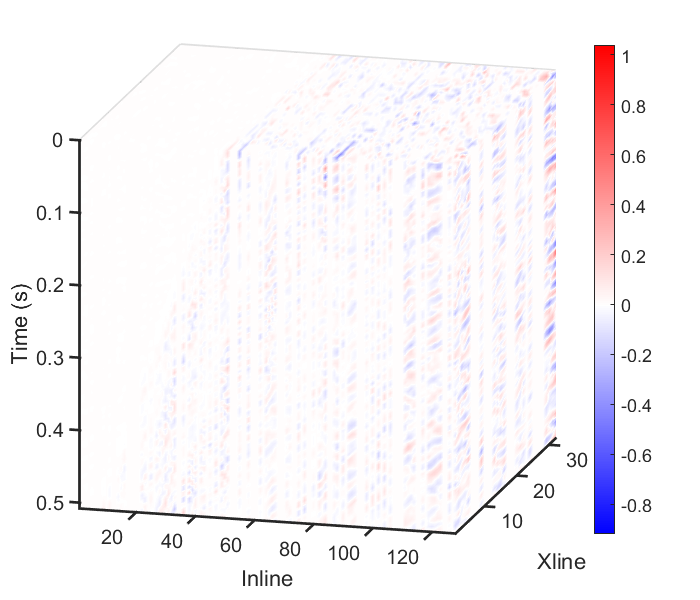}}
	\subfloat[]{\includegraphics[width=0.2\linewidth]{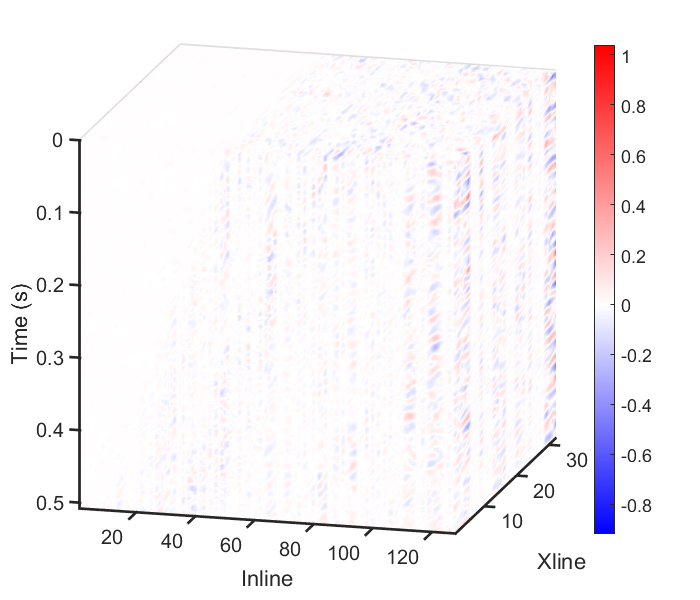}}
	\caption{Reconstruction results of the prestack field data. (a) Ground truth from Mobil Avo Viking Graben Line 12. (b)-(e) Reconstruction results for DRR, DIP, 3-DPCNN, and S2DM, respectively. (f) Ground truth with 50\% randomly missing traces. (g)-(j) Residuals corresponding to DRR, DIP, 3-DPCNN, and S2DM.}
\end{figure}
\begin{figure}[htpb]
	\centering
	\subfloat[]{\includegraphics[width=0.2\linewidth]{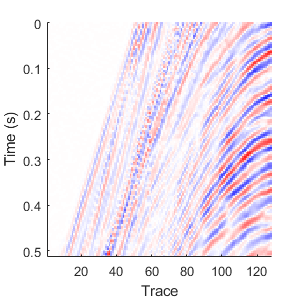}}
	\subfloat[]{\includegraphics[width=0.2\linewidth]{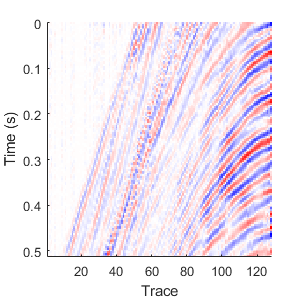}}
	\subfloat[]{\includegraphics[width=0.2\linewidth]{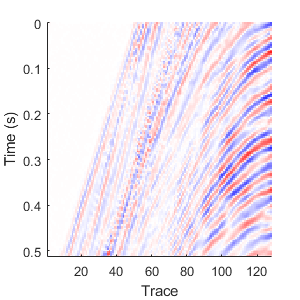}}
	\subfloat[]{\includegraphics[width=0.2\linewidth]{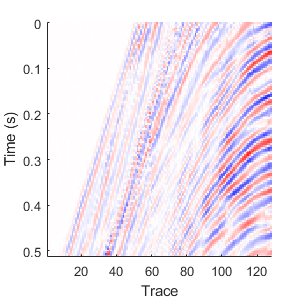}}
	\subfloat[]{\includegraphics[width=0.2\linewidth]{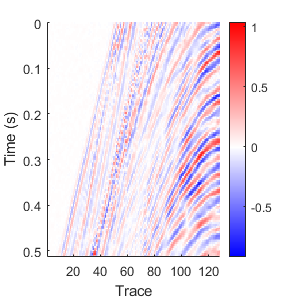}}
	\qquad
	\subfloat[]{\includegraphics[width=0.2\linewidth]{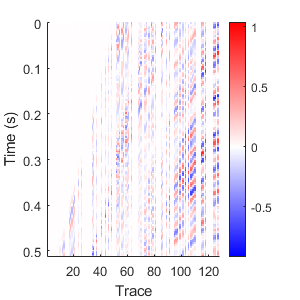}}
	\subfloat[]{\includegraphics[width=0.2\linewidth]{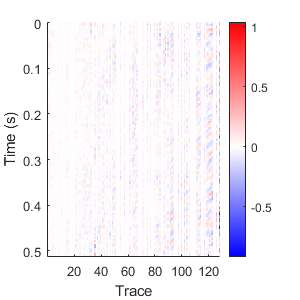}}
	\subfloat[]{\includegraphics[width=0.2\linewidth]{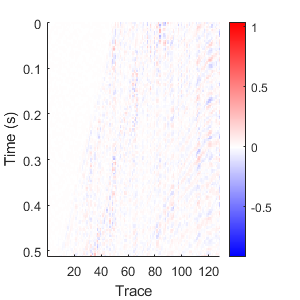}}
	\subfloat[]{\includegraphics[width=0.2\linewidth]{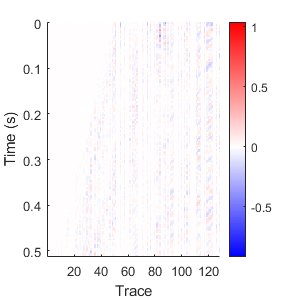}}
	\subfloat[]{\includegraphics[width=0.2\linewidth]{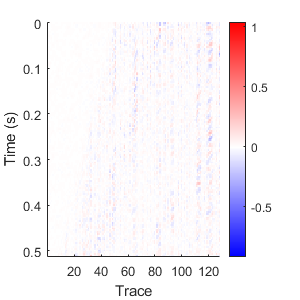}}
	
	\caption{(a)-(j) 2D slices of the data displayed in Figure 10a-10j, respectively.}
\end{figure}
\begin{figure}[htpb]
	\centering
	\subfloat[]{\includegraphics[width=0.2\linewidth]{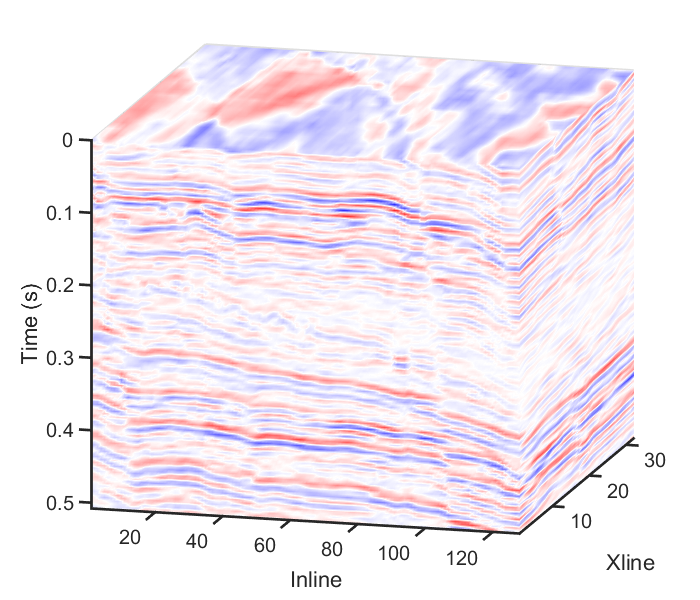}}
	\subfloat[]{\includegraphics[width=0.2\linewidth]{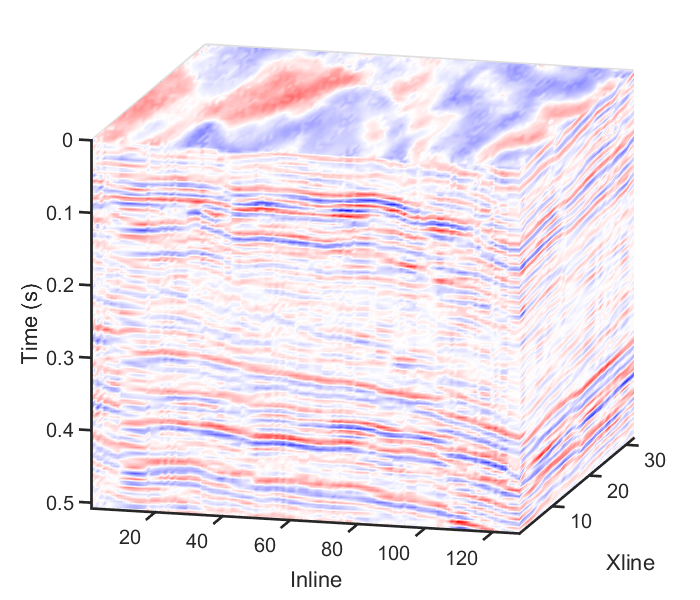}}
	\subfloat[]{\includegraphics[width=0.2\linewidth]{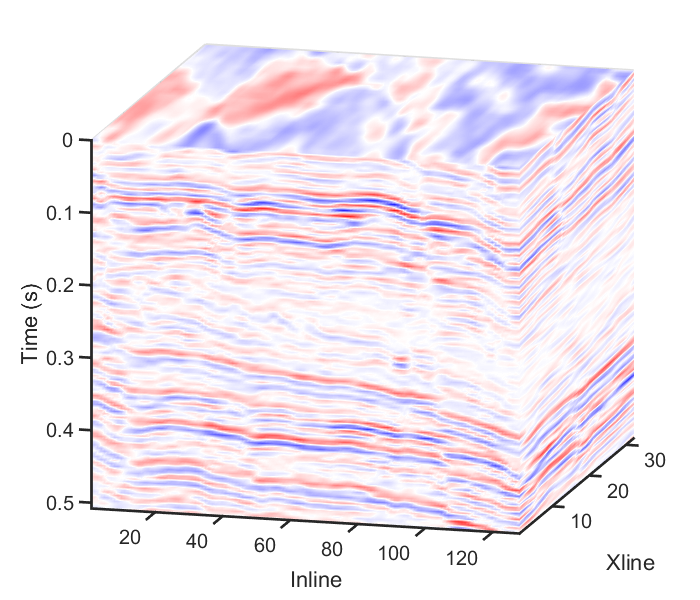}}
	\subfloat[]{\includegraphics[width=0.2\linewidth]{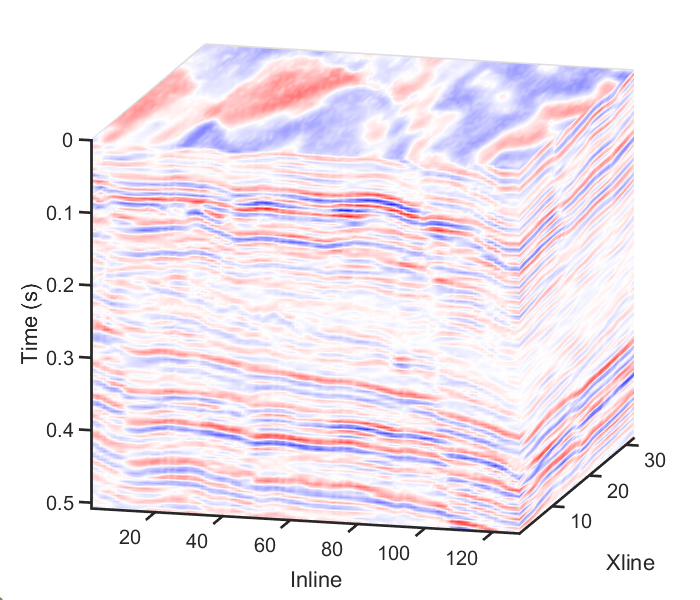}}
	\subfloat[]{\includegraphics[width=0.2\linewidth]{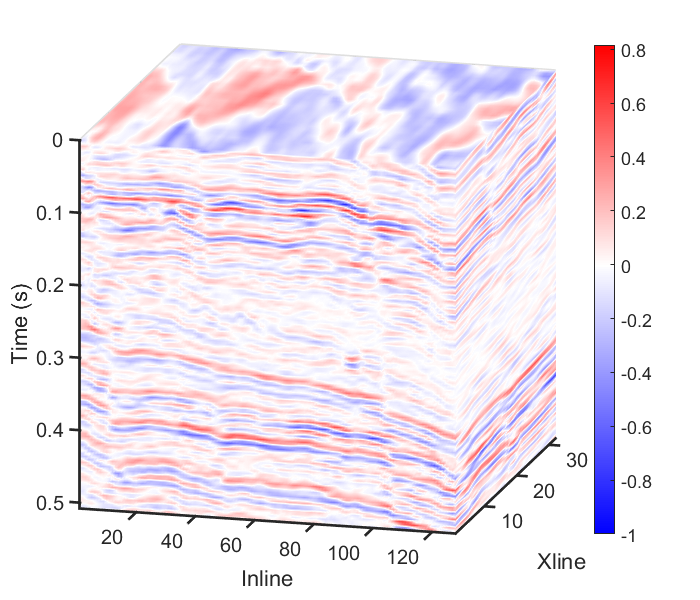}}
	\qquad
	\subfloat[]{\includegraphics[width=0.2\linewidth]{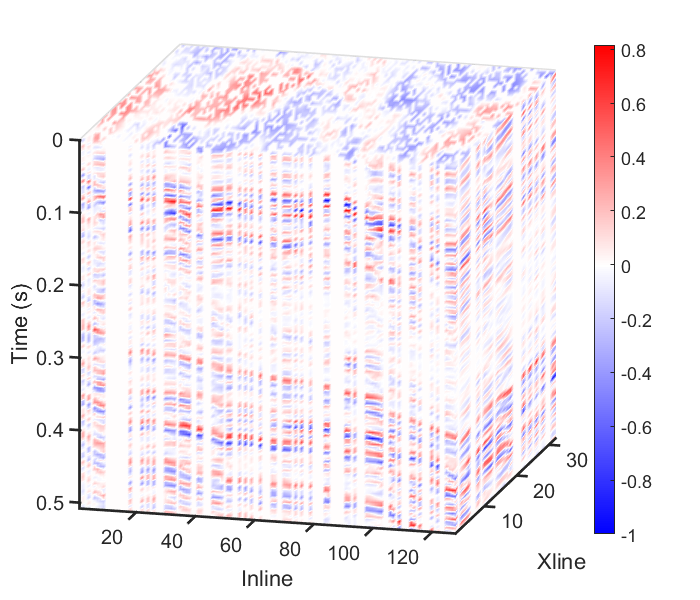}}
	\subfloat[]{\includegraphics[width=0.2\linewidth]{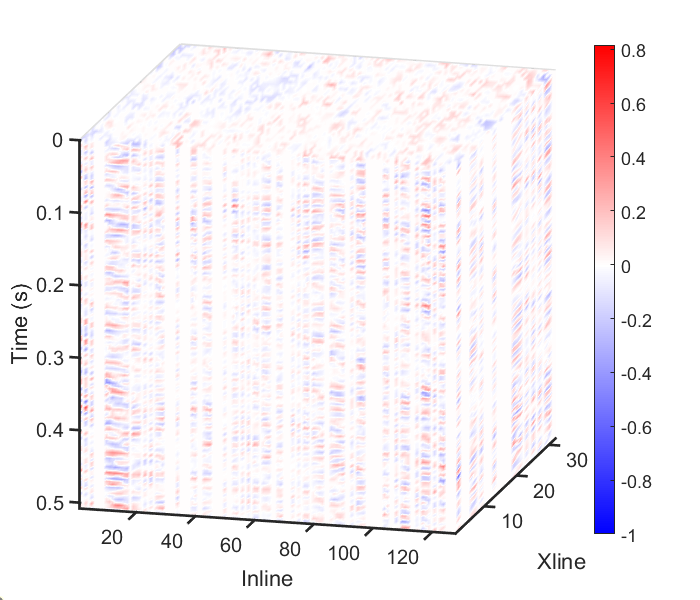}}
	\subfloat[]{\includegraphics[width=0.2\linewidth]{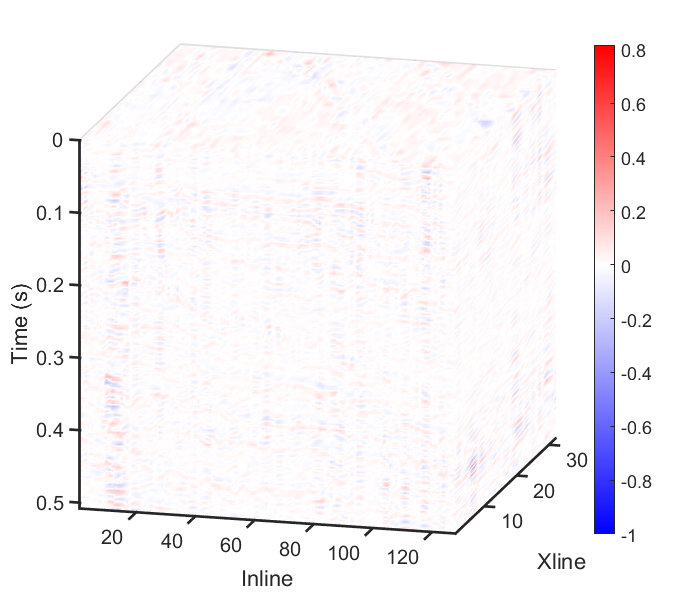}}
	\subfloat[]{\includegraphics[width=0.2\linewidth]{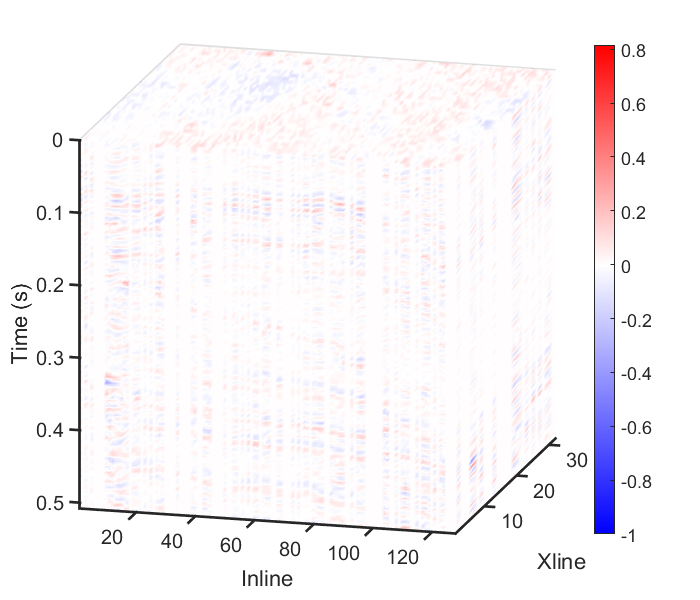}}
	\subfloat[]{\includegraphics[width=0.2\linewidth]{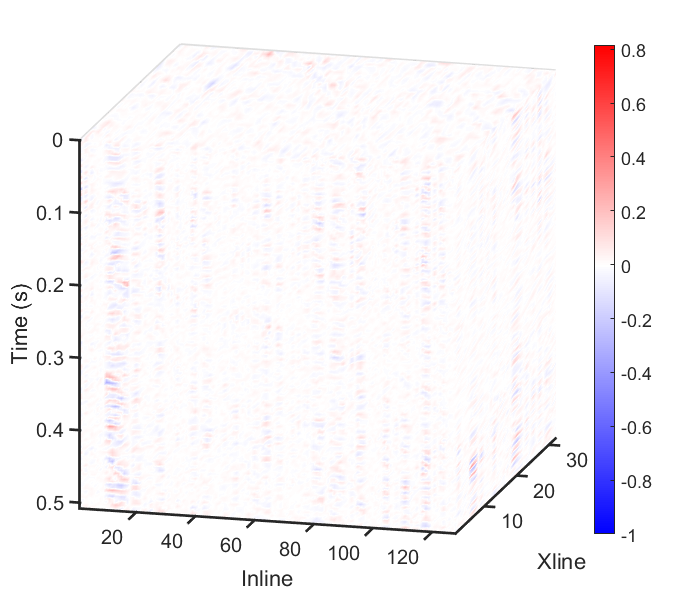}}
	\caption{Reconstruction results of the poststack field data. (a) Ground truth from Kerry-3D. (b)-(e) Reconstruction results for DRR, DIP, 3-DPCNN, and S2DM, respectively. (f) Ground truth with 50\% randomly missing traces. (g)-(j) Residuals corresponding to DRR, DIP, 3-DPCNN, and S2DM.}
\end{figure}
\begin{figure}[htpb]
	\centering
	\subfloat[]{\includegraphics[width=0.2\linewidth]{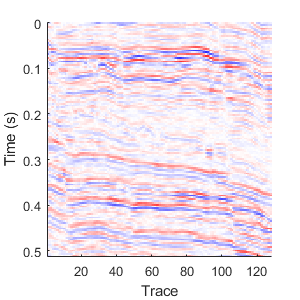}}
	\subfloat[]{\includegraphics[width=0.2\linewidth]{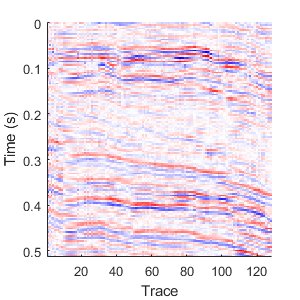}}
	\subfloat[]{\includegraphics[width=0.2\linewidth]{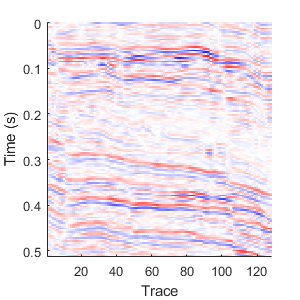}}
	\subfloat[]{\includegraphics[width=0.2\linewidth]{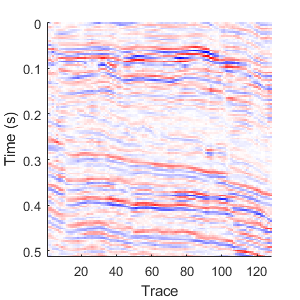}}
	\subfloat[]{\includegraphics[width=0.2\linewidth]{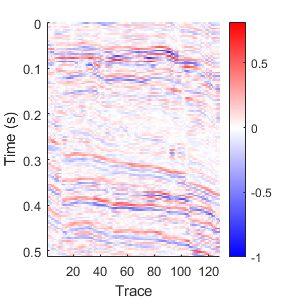}}
	\qquad
	\subfloat[]{\includegraphics[width=0.2\linewidth]{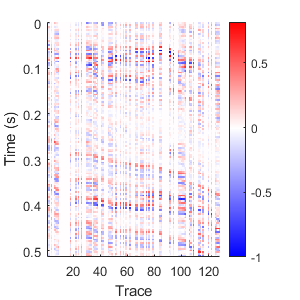}}
	\subfloat[]{\includegraphics[width=0.2\linewidth]{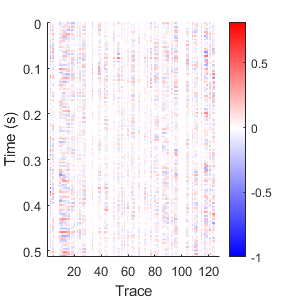}}
	\subfloat[]{\includegraphics[width=0.2\linewidth]{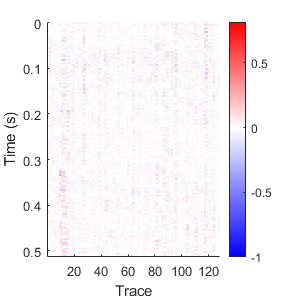}}
	\subfloat[]{\includegraphics[width=0.2\linewidth]{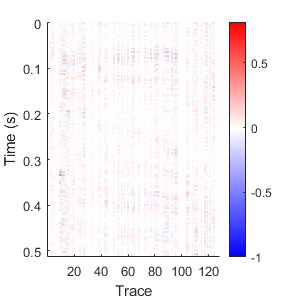}}
	\subfloat[]{\includegraphics[width=0.2\linewidth]{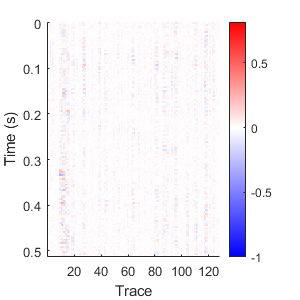}}
	
	\caption{(a)-(j) 2D slices of the data displayed in Figure 12a-12j, respectively.}
\end{figure}

\begin{table}[H]  \caption{\textbf{PSNR and SSIM on 1) SEG C3 and 2) Parihaka-3D.}}%title  
	\centering  
	\begin{tabular}{ccccc} % four columns  
		\toprule[2pt]  %begin the first line  
		{}& PSNR  1)& SSIM 1)&PSNR  2)&SSIM 2)\\  
		\hline %begin the second line  
		DRR&48.9752dB&0.9956&44.6084dB&0.9955\\
		DIP &47.5898dB&0.9947&45.6710dB&0.9943\\
		3-DPCNN&50.0709dB&0.9960&46.8631dB&0.9964\\
		S2DM&\textbf{50.5103}dB&\textbf{0.9961}&\textbf{48.7054dB}&\textbf{0.9974}\\ \bottomrule[2pt] %begin the third line 
	\end{tabular}  
\end{table}

\begin{table}[H]  \caption{\textbf{PSNR and SSIM on 1) Mobil Avo Viking Graben Line 12 and 2) Kerry-3D.}}%title  
	\centering  
	\begin{tabular}{ccccc} % four columns  
		\toprule[2pt]  %begin the first line  
		{}& PSNR  1)& SSIM 1)&PSNR  2)&SSIM 2)\\  
		\hline %begin the second line  
		DRR&35.1610dB&0.9745&32.1710dB&0.9462\\
		DIP &36.5702dB&0.9810&37.1286dB&0.9775\\
		3-DPCNN&37.1964dB&0.9844&35.9801dB&0.9807\\
		S2DM&\textbf{37.4179dB}&\textbf{0.9857}&\textbf{38.6564dB}&\textbf{0.9851}\\ \bottomrule[2pt] %begin the third line 
	\end{tabular}  
\end{table}

\begin{table}[H]  \caption{\textbf{PSNR and SSIM on Parihaka-3D at different missing levels.}}%title  
	\centering  
	\begin{tabular}{cccccc} % four columns  
		\toprule[2pt]  %begin the first line  
		{}& 50\%& 60\%&70\%&80\%&90\%\\  
		\hline %begin the second line  
		\multirow{2}*{DRR}&44.6084dB&42.3601dB&36.8520dB&31.1442dB&25.5198dB\\
		&0.9956&0.9925&0.9690&0.8928&0.6320\\
		\multirow{2}*{DIP}&45.9751dB&44.3247dB&44.0704dB&42.0562dB&39.3059dB\\
		&0.9944&0.9940&0.9936&0.9904&0.9834\\
		\multirow{2}*{3-DPCNN}&46.8632dB&46.0219dB&44.3867dB&42.5243dB&39.6283dB\\
		&0.9966&0.9950&0.9945&0.9915&\textbf{0.9839}\\
		\multirow{2}*{S2DM}&\textbf{48.7054dB}&\textbf{47.2083dB}&\textbf{45.4495dB}&\textbf{42.8757dB}&\textbf{39.7876dB}\\
		&\textbf{0.9974}&\textbf{0.9966}&\textbf{0.9950}&\textbf{0.9919}&\textbf{0.9839}\\
		\bottomrule[2pt] %begin the third line 
	\end{tabular}  
\end{table}

\begin{table}[H]  \caption{\textbf{Computing time comparison. }}%title  
	\centering  
	\begin{tabular}{cccc} % four columns  
		\toprule[2pt]  %begin the first line  
		DRR&DIP&3-DPCNN&S2DM\\  
		\hline %begin the second line  
		0.48h&\textbf{0.21h}&13.74h&2.35h\\ \bottomrule[2pt] %begin the third line 
	\end{tabular}  
\end{table}

\end{document}